\documentclass[journal]{IEEEtran}
\usepackage{setspace}
\usepackage{cite}
\usepackage[bottom]{footmisc}
\usepackage[]{authblk}
\usepackage{graphicx}
\usepackage{epstopdf} 
\usepackage{psfrag}
\usepackage[lofdepth,lotdepth]{subfig}
\usepackage{multicol}
\usepackage{amsmath}
\DeclareMathOperator{\Tr}{Tr}
\usepackage{relsize}
\usepackage{amsfonts}
\usepackage{amsthm}
\usepackage{amssymb}
\usepackage{epstopdf}
\usepackage{color}
\usepackage{xcolor}
\usepackage{tikz}
\usepackage{stfloats}
\usepackage{float}
\usepackage{algorithm}
\usepackage{algpseudocode}
\newtheorem{theorem}{Theorem}

\usetikzlibrary{arrows,shapes,trees} 
\usepackage[justification=centering]{caption}
\makeatletter
\title{Hybrid Precoder and Combiner Designs for Decentralized Parameter Estimation in mmWave MIMO Wireless Sensor Networks}
\author{ Priyanka~Maity,~\IEEEmembership{Student Member,~IEEE,} Suraj~Srivastava,~\IEEEmembership{Member,~IEEE,} Kunwar~Pritiraj~Rajput,~\IEEEmembership{Member,~IEEE,} Naveen~K. D.~Venkategowda,~\IEEEmembership{Member,~IEEE,} Aditya~K.~Jagannatham,~\IEEEmembership{Senior Member,~IEEE,} and Lajos Hanzo,~\IEEEmembership{Life Fellow,~IEEE} %

 \thanks{L. Hanzo would like to acknowledge the financial support of the Engineering and Physical Sciences Research Council projects EP/W016605/1 and EP/X01228X/1 as well as of the European Research Council's Advanced Fellow Grant QuantCom (Grant No. 789028). The work of Aditya K. Jagannatham was supported in part by the
Qualcomm Innovation Fellowship, and in part by the Arun Kumar Chair Professorship.} %
 
  \thanks{P. Maity, S. Srivastava and A. K. Jagannatham are with the
Department of Electrical Engineering, Indian Institute of Technology,
Kanpur, Kanpur, 208016, India (e-mail: pmaity@iitk.ac.in, ssrivast@iitk.ac.in, 
adityaj@iitk.ac.in.)}%

\thanks{K. P. Rajput was with  the Department of Electrical Engineering, Indian Institute of Technology, Kanpur, Kanpur, 208016, India. He is now with the interdisciplinary center of security reliability and trust (SnT), University of Luxembourg, 4365, Luxembourg (e-mail: kunwar.rajput@uni.lu)}%

\thanks{ N. K. D. Venkategowda is with the Department of Science and
Technology, Linköping University,  60174 Norrköping, Sweden (e-mail:
naveen.venkategowda@liu.se.)}%

\thanks{L. Hanzo is with the School of Electronics and Computer
Science, University of Southampton, Southampton SO17 1BJ, U.K.
(e-mail: lh@ecs.soton.ac.uk)}       } 
 
\begin{document}
\maketitle
\begin{abstract}
Hybrid precoder and combiner designs are conceived for decentralized parameter estimation in millimeter wave (mmWave) multiple-input multiple-output (MIMO) wireless sensor networks (WSNs). More explicitly, efficient pre- and post-processing of the sensor observations and received signal are proposed for the minimum mean square error (MMSE) estimation of a parameter vector. The proposed techniques exploit the limited scattering nature of the mmWave MIMO channel for formulating the hybrid transceiver design framework as a multiple measurement vectors (MMV)-based sparse signal recovery problem. This is then solved using the iterative appealingly  low-complexity simultaneous orthogonal matching pursuit (SOMP). Tailor-made designs are presented for WSNs operating under both total and per-sensor power constraints, while considering ideal noiseless as
well as realistic noisy sensors.
 Furthermore, both the Bayesian Cramer-Rao lower bound and the centralized MMSE bound are derived for benchmarking the proposed decentralized estimation schemes. Our simulation results demonstrate the efficiency of the designs advocated and verify the analytical findings.
\end{abstract}

\section{Introduction}
Recent advances in massive Machine-Type Communication (mMTC) and low cost of sensors have opened new opportunities for
deploying large scale wireless sensor
networks (WSNs) \cite{akyildiz2002wireless}. This has ushered in a new era and paved the way  for compelling applications, such as the Internet of Things (IoT) \cite{lin2017survey} \cite{petrov2018iot}.
This in turn enables various applications of substantial socio-economic value, such as health care \cite{pathinarupothi2018iot}, agriculture \cite{delnevo2021deep}, defence \cite{hao2021industrial}, smart cities \cite{venkatesh2017modular}, smart grids \cite{li2017smart}, vehicle-to-everything (V2X) communication \cite{liu2019cooperative}, among others.
In a typical WSN that performs decentralized estimation \cite{xiao2008linear},\cite{behbahani2012linear}, \cite{behbahani2014decentralized} the constituent low-power sensors precode the observations followed by transmission to a fusion center (FC) over a coherent multiple access channel (MAC). The FC subsequently processes the received data for generating reliable estimates of the multiple parameters under consideration. In contrast, centralized estimation refers to the scenario when all the
sensor measurements are available directly at the FC without any distortion. Naturally, the estimation accuracy in WSNs is generally degraded by both the measurement/observation noise at the sensors, as well as by the fading nature of the wireless channel and also by the ubiquitous thermal noise of the communication circuitry at the FC. Therefore, conceiving powerful precoder/combiner designs capable of exploiting the multiplexing and diversity gains of the multiple-input multiple-output (MIMO) WSN is of crucial importance for
achieving reliable estimates of the parameters.
Unfortunately, the massive connectivity required by the large number of sensors in a large-scale IoT deployment further exacerbates the spectrum crunch in the over-crowded sub-6 GHz band. This motivates leveraging the less congested, but higher path loss spectrum beyond the current sub-6 GHz bands.

Millimeter wave (mmWave) wireless technology, which exploits the abundant spectrum available in the mmWave band (30-300 GHz), can help realize the dual goals of massive connectivity and high data rates necessary to support WSNs \cite{sahoo2018enabling}, \cite{chen2018spatial}, \cite{chettri2019comprehensive}. 
However, communication in the mmWave
band is challenging due to its high
propagation losses and severe signal blockage  \cite{rappaport2015millimeter},\cite{heath2016overview}. In such a scenario, multiple antenna technology can be leveraged to overcome
these barriers, which is especially convenient in mmWave
frequency bands owing to the short wavelength that supports
the fabrication and integration of a very large number of
antennas in compact devices. 
However, it must be noted that
the large number of antennas at both ends necessitates a large
number of radio-frequency (RF) chains, since the conventional
baseband transceiver architecture requires an individual RF
chain for each antenna. This leads to unsustainably high power
consumption, especially by the analog-to-digital converters
(ADCs) \cite{zhang2015massive} that are required to operate at high sampling rates due to the
high bandwidth of signals in the mmWave frequency bands.
To overcome this problem, the hybrid RF-baseband transceiver architecture, proposed in the
pioneering contributions \cite{molisch2017hybrid}, \cite{el2014spatially} offers an excellent
solution for the practical realization
of mmWave MIMO systems.
Hence, we focus our attention on designing novel hybrid precoder/combiner designs for decentralized parameter vector estimation in mmWave MIMO WSNs. Next we discuss other novel contributions on linear decentralized estimation.

\subsection{State-of-the-Art in Transceiver Designs for WSNs}
The popular model of linear decentralized estimation has been developed in the seminal works in \cite{xiao2008linear}, \cite{behbahani2012linear} and \cite{6894624}. However, the signal processing techniques proposed therein are unsuitable in mmWave MIMO-based WSNs, given the specific nature of signal processing at mmWave frequencies. Specifically, for mmWave MIMO systems, analog beamforming \cite{heath2016overview} is a compellingly low-complexity, low-power technique. However, its impediment is that it does not readily support multi-stream transmission, which is critical for the simultaneous estimation of multiple parameters. As a remedy, hybrid beamforming techniques have been proposed in \cite{el2014spatially}, which simultaneously support both spatial multiplexing and transmission of multiple streams in mmWave MIMO systems. The hybrid transceiver employed in such systems is comprised of an RF precoder consisting of a digitally controlled network of phase shifters and a baseband precoder relying on a digital signal processor (DSP). Similarly, the hybrid combiner at the receiver is comprised of an analog RF combiner in cascade with a digital baseband combiner. For designing the hybrid transceiver components, Ayach \textit{et al.} \cite{el2014spatially} proposed an interesting hybrid transmit precoder (TPC)/ receive combiner (RC) design for mutual information
maximization in a point-to-point mmWave MIMO system. The conventional fully-digital
optimal precoder obtained therein is the right singular matrix of the mmWave MIMO
channel matrix. The authors then decompose the fully-digital TPC into its RF and baseband
components using the low-complexity simultaneous orthogonal matching pursuit (SOMP) algorithm
incorporating the properties of the mmWave channel model. However, in a coherent
MAC-based mmWave MIMO WSN, the design of the fully-digital TPC is a non-trivial problem. Yu \textit{et al.} \cite{yu2016alternating} formulated the hybrid precoder design paradigm as a matrix
factorization problem, and developed an alternating minimization (Altmin)
algorithm for solving it. The
iterative algorithm presented therein has a significantly higher complexity and does not exploit
the specific properties of the mmWave MIMO channel.
Gong \textit{et al.} \cite{gong2020majorization} proposed hybrid transceiver designs based on majorization-minimization (MM) method.
The other various hybrid precoder and combiner designs for cellular mmWave MIMO systems are described in works such as \cite{lyu2021lattice,chen2020hybrid,feng2020dynamic,cai2020secure,8332507}. 
Furthermore, Wang \textit{et al.} \cite{wang2021robust} addressed the security aspect and designed a hybrid precoder for securing broadcast communications in mmWave IoT networks. Hybrid beamforming was also considered by Chae \textit{et al.} \cite{chae2018simultaneous} for simultaneous wireless information and power transfer (SWIPT) based IoT sensor networks where the nodes were considered working only in half-duplex mode.
\begin{table*}[tbh]
\centering
 \caption{Boldly contrasting our new contributions against the state-of-the-art} \label{table}
\begin{tabular}{||l|c|c|c|c|c|c|c|c|c|c|c|c|c|c|c|c|}
\hline
Attribute  &[20], [21], [24], [25] &[22] &[28]& [29], [30] &[31]&[33] &[34]&Proposed \\
\hline
\hline
mmWave MIMO WSN & &\checkmark &\checkmark & &\checkmark & &\checkmark &\checkmark\\
\hline
Decentralized Estimation  &  & & & \checkmark  &  \checkmark &\checkmark &\checkmark &\checkmark\\
\hline
Coherent MAC  & & & &\checkmark  & &\checkmark &\checkmark &\checkmark\\
\hline

Vector Parameter Estimation& & & &  &\checkmark &\checkmark & &\checkmark \\
\hline
Hybrid Transceiver Design &\checkmark &\checkmark &\checkmark &  &\checkmark &\checkmark & &\checkmark\\
\hline
Closed-form Solution  & \checkmark &\checkmark  &\checkmark & & &\checkmark &\checkmark &\checkmark\\
\hline
MSE-optimal   & &\checkmark & &  &&\checkmark &\checkmark &\checkmark \\
\hline
Total Power Budget    &  & &\checkmark & & & &\checkmark &\checkmark\\
\hline
Individual Sensor Power Constraints  &  & & & \checkmark  & \checkmark &\checkmark &\checkmark &\checkmark\\
\hline
\end{tabular}
\end{table*}  However, authors therein have not exploited the properties
of the mmWave MIMO channel and considered a single antenna at the FC for scalar parameter estimation, hence they were unable to glean any multiplexing gain for improving the estimation accuracy. Zhao \textit{et al.} \cite{zhao2019hybrid} also proposed a hybrid precoding scheme for SWIPT based IoT nodes in a Rayleigh fading channel working in full-duplex mode. Lie \textit{et al.} \cite{liu2020analog} presented a hybrid TPC design based on the alternating direction method of multipliers
(ADMM) for the maximization of mutual
information in an orthogonal MAC. However, the estimation accuracy in a WSN is best quantized in terms of the mean square error (MSE) of parameter estimation, which cannot be represented by a mutual information based objective function. The authors of \cite{4355332} presented a MMSE-based transceiver optimization procedure for
multi-user MIMO systems, where the optimal TPCs are obtained by minimizing the
sum MSE subject to a constraint on the total power of all the users. However, it
must be noted that their scheme cannot be used in WSNs since the system model of a
MIMO WSN is substantially different to that of a multi user MIMO cellular system.
In an uplink multi-user MIMO system, each user transmits its data symbol to the base
station, which is independent of the data symbols of the other users. Therefore, the
data transmitted from the kth user suffers from interference arising due to the other
co-channel users. By contrast, in a WSN, all the sensors simultaneously observe correlated
versions of the same parameter. Thus, the observation vectors transmitted
from the sensors are correlated and noisy versions of the parameter have to be estimated. Moreover, it can be readily inferred from \cite{4355332} that in a multi-user MIMO system, the TPCs $\left\lbrace\overline{\mathbf{T}}_k\right\rbrace_{k=1}^K$ corresponding to the $K$ users are decoupled in the sum-MSE expression. However, in the corresponding sum-MSE expression of the MIMO WSN, the TPCs $\left\lbrace{\mathbf{F}}_k\right\rbrace_{k=1}^K$ corresponding to the $K$ sensors are coupled, as will be seen in Section \ref{SectionIV}. This necessitates a completely different algorithm for the design of the TPCs, which is described
in our paper. Authors in \cite{liu2021hybrid} have proposed ADMM-based iterative hybrid transceiver designs in MIMO WSNs. However, the work does not exploit the mmWave channel sparsity for hybrid transceiver designs. Rajput \textit{et al.} \cite{rajput2021hybrid} have proposed hybrid precoder designs for scalar parameter estimation in mmWave MIMO WSNs. However, to the best of our knowledge, none of the existing contributions have developed a hybrid transceiver design  taking into account the associated power constraints of the parameter vector estimation problem of a mmWave hybrid MIMO WSN, which is hence the objective of this treatise. The contributions of this work are described in the next section.
Furthermore, the novel aspects of this paper are boldly and explicitly contrasted to the existing literature in Table \ref{table}.
\subsection{Our Contributions}
\begin{itemize}
\item Novel TPC/ RC designs are developed for decentralized parameter vector estimation in 
mmWave MIMO WSNs. Our hybrid analog-digital transceiver designs are first designed for a scenario associated with noisy measurements under a given total power budget. Subsequently, hybrid TPCs are designed for a mmWave hybrid MIMO WSN subject to individual power constraints at the constituent sensors, which are suitable for applications where each sensor has a stringent power budget due to its limited battery charge. 
\item  Closed-form expressions are derived for the optimal fully-digital TPCs using majorizaton theory. Subsequently, the hybrid baseband and RF TPC components are designed using SOMP. 
\item For parameter estimation at the FC, both a fully-digital RC, and its baseband and RF components are derived by minimizing the MSE using the classic linear minimum mean square error (LMMSE) framework.
\item Both, the Bayesian Cramer-Rao bound (BCRB) and the centralized MMSE bound
are derived for benchmarking the MSE performance of parameter estimation in the
proposed designs relying on noiseless and noisy sensor observations, respectively. 
\item Our simulation results confirm
that the proposed schemes perform close to the fully-digital designs having as many RF chains as the number of antennas, and they approach the
centralized MMSE lower bound at high signal-to-noise ratio (SNR), hence
illustrating their efficiency.
\end{itemize}     
\subsection{Outline of the paper}
The rest of the paper is outlined as follows. Section II discusses the mmWave hybrid MIMO WSN system model and the mmWave MIMO channel model of linear decentralized parameter estimation. Section III presents MSE-optimal hybrid TPC/ RC designs, initially under a total power budget and individual sensor power constraints. Our simulation results are discussed in Section V, and we conclude in Section VI. For maintaining seamless flow, the proofs of the various propositions are relegated to the Appendices. \par
\textbf{Notation:} Small and capital boldface letters $\mathbf{z}$ and $\mathbf{Z}$ have been used to represent vectors and matrices, respectively. $()^T$, $()^H$ and $()^{\dagger}$ denote transpose, Hermitian transpose and pseudo-inverse operations respectively; Furthermore, the $(i,j)$th element of a matrix $\mathbf{Z}$ is denoted by $\mathbf{Z}(i,j)$; A block diagonal matrix is denoted by $\mathrm{blkdiag}(\mathbf{Z}_1,\mathbf{Z}_2,\hdots,\mathbf{Z}_L)$ where the matrices $\mathbf{Z}_1$, $\mathbf{Z}_2$ upto $\mathbf{Z}_L$ are on its principal diagonal; Furthermore, $\mathbf{\mathrm{diag}\left(\mathbf{Z}\right)}$ and $\mathbf{\boldsymbol{\lambda}\left(\mathbf{Z}\right)}$ denote the vector of diagonal elements and vector of eigenvalues of a matrix $\mathbf{Z}$ respectively; $\mathbf{{\lambda}}_k\left(\mathbf{Z}\right)$ and $\mathbf{{\sigma}}_k\left(\mathbf{Z}\right)$ represent the $k$th eigenvalue and $k$th singular value of the matrix $\mathbf{Z}$, respectively; The column-space and row-space of matrix $\mathbf{Z}$ are denoted by $\mathcal{C}(\mathbf{Z})$ and  $\mathcal{R}(\mathbf{Z})$, respectively;
The operator $\left(z\right)^{+}$ denotes $\text{max}\left\lbrace z,0\right\rbrace$; $\mathcal{CN}(0,1)$ denotes a standard complex normal distributed random variable with mean $0$ and variance $1$. 
\begin{figure*}
\center
{\includegraphics[scale=0.5]{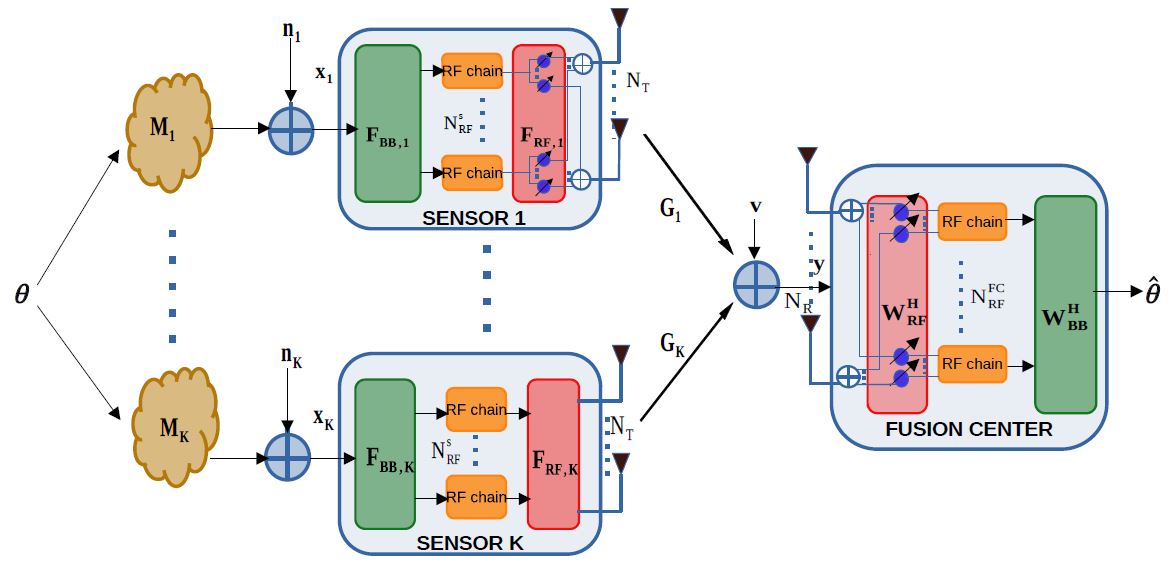}}
\caption{Coherent MAC based mmWave MIMO WSN system model}
\label{fig:sys_mod}
\end{figure*}
\section{mmWave Hybrid MIMO WSN System Model}
This section describes the mmWave hybrid MIMO WSN system model and the mmWave MIMO channel model of linear decentralized estimation.
\subsection{mmWave MIMO WSN Model}
The mmWave WSN, as depicted in Fig. \ref{fig:sys_mod}, consists of $K$ sensors each having $N_T$ antennas, and $N_{\text{RF}}^{s}$  RF chains and a FC equipped with $N_R$ antennas and $N_{\text{RF}}^{\text{FC}}$ RF chains, where $N_{\text{RF}}^{\text{FC}}\ll N_R$ and $ N_{\text{RF}}^s\ll N_T $. Each sensor simultaneously senses the common parameter vector $\boldsymbol{\theta} \in {\mathbb{C}}^{m \times 1}$, where $\boldsymbol{\theta} \sim \mathcal{CN}(\mathbf{0},\mathbf{R}_{\theta})$. The observations ${\mathbf{x}}_k \in {\mathbb{C}}^{q_k \times 1}$ at the $k$th sensor can be modelled as
\begin{align} \label{eq:BB}
{\mathbf{x}}_k={\mathbf{M}}_k\boldsymbol{\theta}+{\mathbf{n}}_k.
\end{align}
The quantity ${\mathbf{M}}_k\in {\mathbb{C}}^{q_k\times m}$ denotes the measurement matrix of the $k$th sensor, while $q_k$ represents the number of measurements sensed by the $k$th sensor. Here ${\mathbf{n}}_k\in {\mathbb{C}}^{q_k \times 1}$ represents the circularly symmetric complex additive Gaussian observation noise at the $k$th sensor with zero mean and covariance matrix $\mathbf{R}_{n,k} \in \mathbb{C}^{q_k \times q_k}$.

As per the hybrid TPC scheme, the observation vector is precoded initially by the baseband TPC $\mathbf{F}_{\text{BB},k} \in {\mathbb{C}}^{N_{\text{RF}}^s \times q_k}$ followed with the RF TPC $\mathbf{F}_{\text{RF},k} \in {\mathbb{C}}^{N_T\times N_{\text{RF}}^s}$ of the $k$th sensor. The RF TPC is realized exclusively using
equal-gain phase shifters, and performs analog processing in the RF domain to achieve a beamforming gain. Therefore, the magnitudes of the elements of RF TPC can be constrained as $\vert \mathbf{F}_{\text{RF},k}(s,t)\vert = \frac{1}{\sqrt{N_T}} $. The mmWave MIMO channel between the $k$th sensor and the FC is denoted by ${ \mathbf{G} }_{ k } \in {\mathbb{C}}^{N_R \times N_T} $. Thus, the signal received by the FC can be expressed as 
\begin{align} \label{eq6}
{ \mathbf{y} } & =\sum_{k=1}^{K} { \mathbf{G} }_{ k }{\mathbf{F}}_{ \text{RF},k }{ \mathbf{F} }_{ \text{BB},k }{ \mathbf{x} }_{ k }+{ \mathbf{v} } \nonumber \\
&=  \sum_{k=1}^{K} { \mathbf{G} }_{ k }{\mathbf{F}}_{ \text{RF},k }{ \mathbf{F} }_{ \text{BB},k }{ \mathbf{M} }_{ k }\boldsymbol{\theta} +\sum_{k=1}^{K}{ \mathbf{G} }_{ k }{\mathbf{F}}_{\text{RF},k }{ \mathbf{F} }_{ \text{BB},k }{ \mathbf{n} }_{ k } +{ \mathbf{v} }\nonumber \\
&= { \mathbf{G} }{\mathbf{F}}{ \mathbf{M} } \boldsymbol{\theta} + { \mathbf{G} }{\mathbf{F}}\mathbf{n}+ { \mathbf{v} },
\end{align}
where $\mathbf{v} \in {\mathbb{C}}^{N_R\times 1}$ is the FC noise, which is distributed as $\mathbf{v} \sim \mathcal{CN}(\mathbf{0},\mathbf{R}_v)$, where  $\mathbf{R}_v =\sigma_v^2\mathbf{I}_{N_R}$. The stacked observation matrix $\mathbf{M} \in {\mathbb{C}}^{q \times m}$, and the concatenated channel matrix $\mathbf{G} \in {\mathbb{C}}^{N_R\times KN_T}$ across the sensors, are defined as
\begin{align}
\mathbf{M} &= \left[\mathbf{M}_1^T,\mathbf{M}_2^T,\cdots,\mathbf{M}_K^T\right]^T,\mathbf{G} = \left[\mathbf{G}_1,\mathbf{G}_2,\cdots,\mathbf{G}_K\right],
\end{align}
where $\sum_{k=1}^{K}q_k=q$. The block-diagonal hybrid TPC $\mathbf{F} \in {\mathbb{C}}^{KN_T\times q}$ is given by
\begin{equation}
\mathbf F = \mathbf{F}_{\text{RF}}\mathbf{F}_{\text{BB}},
\end{equation} 
where ${\mathbf{F}_{\text{RF}}} \in \mathbb{C}^{KN_T \times KN_{\text{RF}}^{s}}$ is the block-diagonal RF TPC and $\mathbf{F}_{\text{BB}}\in \mathbb{C}^{KN_{\text{RF}}^{s} \times q}$ is the block-diagonal baseband TPC, which are defined as 
\begin{align}
\mathbf{F}_{\text{RF}}=\mathrm{blkdiag} (\mathbf{F}_{\text{RF},1} ,\mathbf{F}_{\text{RF},2} , \cdots ,\mathbf{F}_{\text{\text{RF}},K}), \\
\mathbf{F}_{\text{BB}}= \mathrm{blkdiag} ( \mathbf{F}_{\text{BB},1},  \mathbf{F}_{\text{BB},2}, \cdots , \mathbf{F}_{\text{BB},K}).
\end{align}
Furthermore, the covariance matrix $\mathbf{R}_n\in {\mathbb{C}}^{q \times q}$ of the concatenated observation noise vector $\mathbf{n} = [\mathbf{n}_1^T, \mathbf{n}_2^T, \cdots ,\mathbf{n}_K^T]^T \in \mathbb{C}^{q \times 1}$ is given by
\begin{align}
&\mathbf{R}_n= \mathrm{blkdiag} (\mathbf{R}_{n,1}, \mathbf{R}_{n,2}, \cdots ,\mathbf{R}_{n,K}). 
\end{align}
\begin{figure}[t]
\centering
  \includegraphics[width=0.45\textwidth]{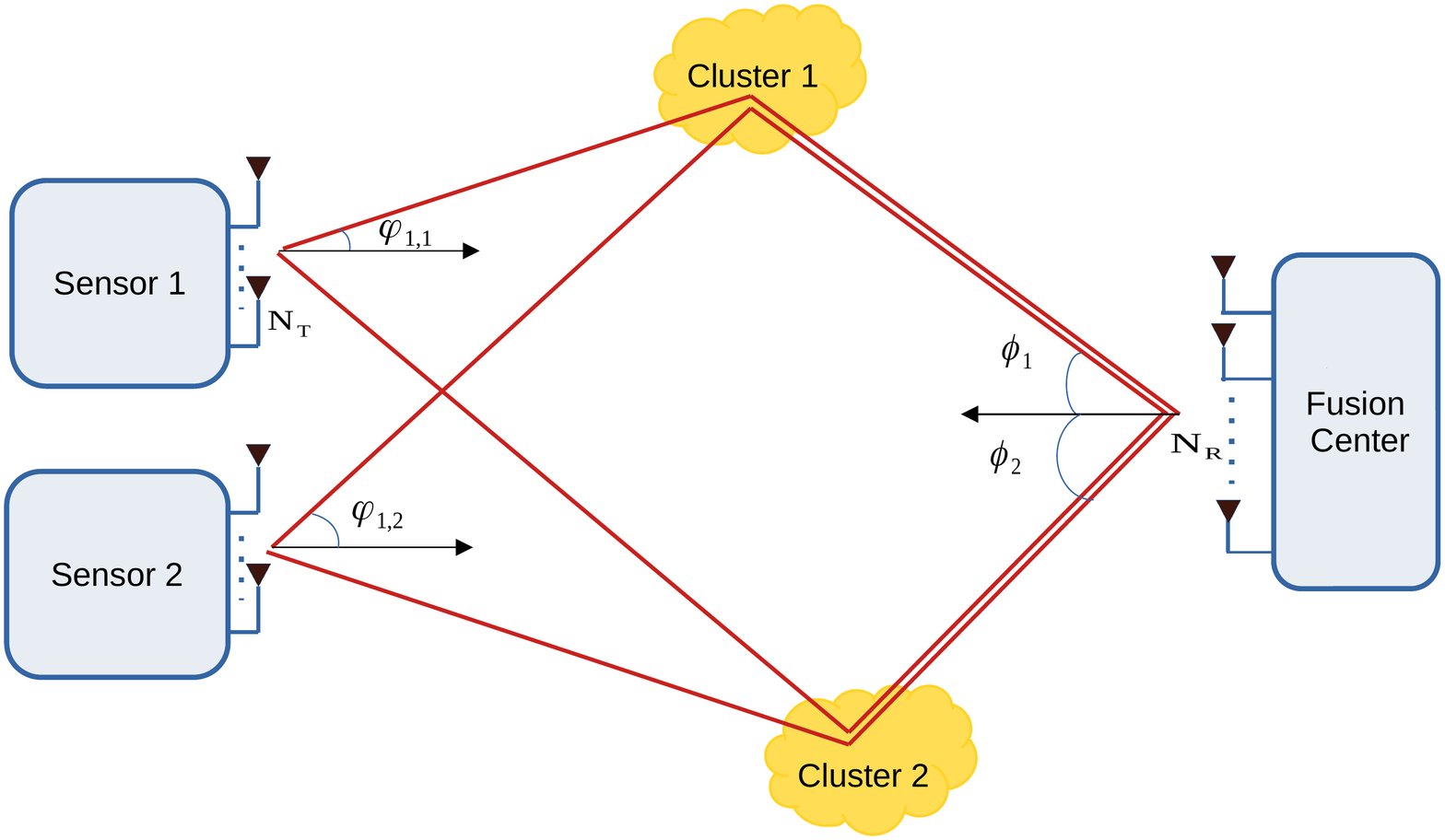}
  \caption{Spatial channel model for the mmWave MIMO system}
  \label{fig:test2}
\end{figure} 
 \begin{figure*}[b]
 \hrulefill
\begin{align}
\widehat{\boldsymbol\theta} &= \mathbf{W}_{\text{BB}}^H \mathbf{W}_{\text{RF}}^H \Bigg(\sum _{ k=1 }^{ K }{ { \mathbf{G} }_{ k } \mathbf{F}_{\text{RF},k} \mathbf{F}_{\text{BB},k}  { \mathbf{M} }_ k }\Bigg)\boldsymbol{\theta} +\mathbf{W}_{\text{BB}}^H \mathbf{W}_{\text{RF}}^H \left( \sum _{ k=1 }^{ K }{ { \mathbf{G} }_{ k } \mathbf{F}_{\text{RF},k} \mathbf{F}_{\text{BB},k} { \mathbf{n} }_{ k } }\right) +\mathbf{W}_{\text{BB}}^H \mathbf{W}_{\text{RF}}^H\mathbf{v} \nonumber \\
 & = \mathbf{W}_{\text{BB}}^H \mathbf{W}_{\text{RF}}^H { \mathbf{G} }{ \mathbf{F} }{ \mathbf{M} } \boldsymbol{\theta} + \mathbf{W}_{\text{BB}}^H \mathbf{W}_{\text{RF}}^H { \mathbf{G} }{ \mathbf{F} }\mathbf{n} + \mathbf{W}_{\text{BB}}^H \mathbf{W}_{\text{RF}}^H\mathbf{v} = \mathbf{W}^H{ \mathbf{G} }{ \mathbf{F} }{ \mathbf{M} } \boldsymbol{\theta} + \mathbf{W}^H { \mathbf{G} }{ \mathbf{F} }\mathbf{n} + \mathbf{W}^H \mathbf{v}.  \label{eq7}
\end{align}
\end{figure*}
The parameter estimate $\widehat{\boldsymbol\theta} \in {\mathbb{C}}^{m\times 1}$ at the FC of the unknown parameter $\boldsymbol\theta$ is obtained using the hybrid RC denoted by $\mathbf{W}$, which can be decomposed as $\mathbf{W}=\mathbf{W}_{\text{RF}}\mathbf{W}_{\text{BB}}$, where $\mathbf{W}_{\text{RF}} \in \mathbb{C}^{N_R \times N_{\text{RF}}^{\text{FC}}} $ is the RF RC and $\mathbf{W}_{\text{BB}} \in \mathbb{C}^{N_{\text{RF}}^{\text{FC}} \times m}$ is the baseband RC. The equivalent system model can be compactly written as shown in \eqref{eq7}. The covariance matrix $\mathbf{E} \in {\mathbb{C}}^{m \times m} $ of the estimation error of parameter $\boldsymbol{\theta}$ is defined as 
\begin{eqnarray} \label{E3}
{\mathbf{E}} & = & \mathbb{E}\left \{ \left ( \widehat{\boldsymbol{\theta}}-\boldsymbol{\theta} \right )\left ( \widehat{\boldsymbol{\theta}}-\boldsymbol{\theta} \right ) ^H\right \},
\end{eqnarray}
whereas the MSE at the FC is given by 
\begin{equation} \label{E4}
\text{MSE}=\text{Tr}\left(\mathbf{E}\right).
\end{equation}
Furthermore, the total transmit power of all the sensors may be formulated as 
\begin{align}\label{eq9}
\mathbb{E}\left\lbrace\vert\vert\mathbf{Fx}\vert\vert_2^2 \right\rbrace =\text{Tr}\left( \mathbf{F} \left( \mathbf{M} \mathbf{M}^H + \mathbf{R}_n \right) { \mathbf{F} }^{ H }\right).
 \end{align}
The relevant mmWave MIMO
channel model is presented in the next subsection.
\subsection{mmWave  MIMO Channel Model}
The mmWave MIMO channel ${\mathbf{G}_k}$ between the FC and the $k$th sensor is modelled using the spatial channel model of \cite{alkhateeb2014channel} having $L$ multipath components formulated as
\begin{align} \label{E7}
\mathbf{G}_k &= \sqrt{\frac{N_RN_T}{L}} \sum _{ n=1 }^{ L } \alpha_{n,k} \boldsymbol{\mathrm{a}}_{R}(\phi_{n})\boldsymbol{\mathrm{a}}_{T}^H(\varphi_{n,k}
), 
\end{align}
where the 3-tuple $(\alpha_{n,k},\phi_{n},\varphi_{n,k})$ denotes the complex gain $\alpha_{n,k}$, the angle of arrival (AoA) $\phi_{n}$ at the FC and angle of departure (AoD) $\varphi_{n,k}$ at the $k$th sensor associated with the $n$th cluster. The spatial channel model of the  mmWave MIMO WSN is presented in Fig. \ref{fig:test2}.
Upon considering uniform linear arrays (ULAs) at the receiver and transmitters, the receive array response vector $\boldsymbol{\mathrm{a}}_{R}(\phi_{n}) \in \mathbb{C}^{N_R \times 1}$ and transmit array response vectors $\boldsymbol{\mathrm{a}}_{T}(\varphi_{n,k}) \in \mathbb{C}^{N_T \times 1}$ are given by
\begin{align}
\boldsymbol{\mathrm{a}}_{R}(\phi_{n})&=\frac{1}{\sqrt{N_R}}\left[ 1,e^{-j\widetilde{\phi}_{n}},...,e^{-j(N_R-1)\widetilde{\phi}_{n}}\right]^T,\\
\boldsymbol{\mathrm{a}}_{T}(\varphi_{n,k})&=\frac{1}{\sqrt{N_T}}\left[ 1,e^{-j\widetilde{\varphi}_{n,k}},...,e^{-j(N_T-1)\widetilde{\varphi}_{n,k}}\right]^T,
\end{align}
where $\widetilde{\phi}_{n}=\frac{2\pi}{\lambda}d_R\text{cos}(\phi_{n})$ and $\widetilde{\varphi}_{n,k}=\frac{2\pi}{\lambda}d_T\text{cos}(\varphi_{n,k})$. The quantities $\lambda,d_R$ and $d_T$ denote the carrier wavelength, and the inter-antenna  spacings at the FC and each sensor, respectively. The mmWave MIMO channel $\mathbf{G}_k$ can now be expressed in the compact form of:
\begin{equation} \label{E10}
\mathbf{G}_k= \mathbf{A}_{R}\mathbf{D}_k\mathbf{A}_{T,k}^H,
\end{equation}
where $\mathbf{D}_k = \sqrt{\frac{N_RN_T}{L}} \mathrm{diag}(\alpha_{1,k},\alpha_{2,k},...,\alpha_{L,k}) \in \mathbb{C}^{L \times L}$ is a diagonal matrix containing the path gains of the channel on its principal diagonal. The quantities $\mathbf{A}_{T,k} \in \mathbb{C}^{N_T \times L}$, $\mathbf{A}_{R} \in \mathbb{C}^{N_R \times L}$ denote the transmit and receive array response matrices, respectively, which are given by
\begin{align}
&\mathbf{A}_{T,k} = \left[\boldsymbol{\mathrm{a}}_{T}(\varphi_{1,k}),\boldsymbol{\mathrm{a}}_{T}(\varphi_{2,k}),...,\boldsymbol{\mathrm{a}}_{T}(\varphi_{L,k})\right], \\
&\mathbf{A}_{R} = \left[\boldsymbol{\mathrm{a}}_{R}(\phi_{1}),\boldsymbol{\mathrm{a}}_{R}(\phi_{2}),...,\boldsymbol{\mathrm{a}}_{R}(\phi_{L})\right].
\end{align}
The concatenated mmWave channel can now be represented as 
\begin{align} \label{E11}
 \mathbf{G} &= [\mathbf{G}_1,\mathbf{G}_2,\cdots,\mathbf{G}_K] \nonumber \\
 &=\mathbf{A}_{R}[\mathbf{D}_{1}\mathbf{A}_{T,1}^H,\mathbf{D}_{2}\mathbf{A}_{T,2}^H,\cdots,\mathbf{D}_{K}\mathbf{A}_{T,K}^H] \nonumber \\
 &= \mathbf{A}_{R}{\mathbf{D}}\widetilde{\mathbf{A}}_{T}^H,
\end{align}
where ${\mathbf{D}}=[\mathbf{D}_{1},\mathbf{D}_{2},\cdots,\mathbf{D}_{K}] \in \mathbb{C}^{L \times KL}$ and $ \widetilde{\mathbf{A}}_{T}^H=\text{blkdiag}\left(\mathbf{A}_{T,1}^H,\mathbf{A}_{T,2}^H,\cdots,\mathbf{A}_{T,K}^H\right) \in \mathbb{C}^{KL \times KN_t}$.
The optimization problem of the TPC sensor matrices $\mathbf{F}_k = \mathbf{F}_{\text{RF},k} \mathbf{F}_{\text{BB},k}\  \forall k $ can now be formulated by minimizing the MSE under a given power constraint. The design procedures of various scenarios are described next.

\section{Hybrid MMSE Precoder/Combiner Design} \label{SectionIV} 
In this section, our hybrid TPC is designed for both total power budget as well as per sensor power constraints and hybrid RC design of the FC. 
\subsection{TPC design under total power budget} \label{SectionIVA}
For the system having a total power budget across all the sensors, the fully-digital LMMSE RC $ \mathbf{W}\in \mathbb{C}^{N_R \times m}$ used for estimating the underlying parameter $\boldsymbol{\theta}$ subject to noisy sensor observations is given by \cite{kay1993fundamentals},
\begin{align}
\mathbf{W}&=\left(\mathbb{E}\left\{\mathbf{y}\mathbf{y}^H   \right\}\right)^{-1}\mathbb{E}\left\{\mathbf{y}\boldsymbol{\theta}^H \right\} \nonumber \\
&=\left( {\mathbf{G}}\mathbf{F}\left(\mathbf{M}{\mathbf{M}}^H+\mathbf{R}_n\right)\mathbf{F}^H\mathbf{G}^H + \sigma_v^2\mathbf{I}_{N_R}\right)^{-1} \mathbf{G}\mathbf{F} {\mathbf{M}} \nonumber \\
&=\mathbf{G}\mathbf{F} \left( \left(\mathbf{M}{\mathbf{M}}^H+\mathbf{R}_n\right)\mathbf{F}^H\mathbf{G}^H{\mathbf{G}}\mathbf{F} + \sigma_v^2\mathbf{I}_{q}\right)^{-1} {\mathbf{M}},  \label{eq44}
\end{align}
where the last step follows using the matrix identity $\left(\mathbf{I}+\mathbf{UAV}\right)^{-1}\mathbf{U}=\mathbf{U}\left(\mathbf{I}+\mathbf{AVU}\right)^{-1}$. The resultant error covariance matrix is shown in \eqref{E38}.
\begin{figure*}[t]
\begin{equation} \label{E38}
\mathbf{E}=\left( \mathbf{I}_m + \mathbf{M}^H\mathbf{F}^H\mathbf{G}^H \left(\mathbf{G}\mathbf{F}\mathbf{R}_n\mathbf{F}^H\mathbf{G}^H+\sigma_v^2\mathbf{I}_{N_R}\right) ^{-1}\mathbf{G}\mathbf{F}\mathbf{M}\right )^{-1}.
\end{equation}
\hrulefill
\end{figure*}
\begin{figure*}[t]
\begin{align} 
\mathbf{E}&=\left( \mathbf{I}_m + 
\mathbf{M}^H\mathbf{R}_n^{-1}\mathbf{M}-\mathbf{M}^H\mathbf{R}_n^{-1}\left(\frac{1}{\sigma_v^2}\mathbf{F}^H\mathbf{G}^H\mathbf{G}\mathbf{F}+\mathbf{R}_n^{-1}\right) ^{-1}\mathbf{R}_n^{-1}\mathbf{M}\right )^{-1} \nonumber\\
&=\left( \mathbf{I}_m + 
\frac{1}{\sigma_n^2}\mathbf{M}^H\mathbf{M}-\frac{1}{\sigma_n^2}\mathbf{M}^H\left(\frac{\sigma_n^2}{\sigma_v^2}\mathbf{F}^H\mathbf{G}^H\mathbf{G}\mathbf{F}+\mathbf{I}_{q}\right) ^{-1}\mathbf{M}\right )^{-1}. \label{err_cov_noisy}
\end{align}
\hrulefill
\end{figure*}
Employing the Woodbury matrix identity  \cite{henderson1981deriving} for simplifying the term $\left(\mathbf{G}\mathbf{F}\mathbf{R}_n\mathbf{F}^H\mathbf{G}^H+\sigma_v^2\mathbf{I}_{N_R}\right) ^{-1}$ and considering the noise covariance matrix $\mathbf{R}_n=\sigma_n^2\mathbf{I}_{q}$, the error covariance matrix can be formulated as shown in \eqref{err_cov_noisy}. Our hybrid TPC design based on MSE minimization can be
formulated as

\begin{equation} 
\begin{aligned} \label{eq5}
& \mathop{\text{minimize}}_{\left\lbrace \mathbf{F}_{\text{RF},k},\mathbf{F}_{\text{BB},k}\right\rbrace _{k=1}^K}
\  \text{Tr}\left(\mathbf{E}\right)  \nonumber \\
& \quad \  \text{subject to}  \quad
\  \textrm{Tr} \left( \mathbf{F}_{\text{RF}}\mathbf{F}_{\text{BB}} \left( \mathbf{M} \mathbf{M}^H + \sigma_n^2\mathbf{I}_{q} \right){\mathbf{F}_{\text{BB}}}^{ H }{ \mathbf{F}_{\text{RF}}}^{ H }\right) \le {P}_{T}, \nonumber \\
&\quad \quad \quad \quad \quad \quad \ \ \vert \mathbf{F}_{\text{RF},k}(s,t)\vert = \frac{1}{\sqrt{N_T}}, \  1 \leq k \leq K,
\end{aligned}
\end{equation}
where $P_T$ denotes the WSN's total transmit power budget. The aforementioned problem is non-convex and intractable owing to the constant modulus constraint of the RF TPC $\mathbf{F}_{\text{RF},k}$. To circumvent this, we first design the optimal fully-digital  hybrid TPC given by $\mathbf{F}=\mathbf{F}_{\text{RF}}\mathbf{F}_{\text{BB}}$. The resultant hybrid TPC is then decomposed into its baseband and RF components. The convex optimization problem of the fully-digital TPC $\mathbf{F}$ is thus formulated as follows
\begin{equation} \label{ee1}
\begin{aligned}
& \underset{\mathbf{{F}}}{\text{minimize}}
& &   \text{Tr}\left( \mathbf{E}\right)    \\
&\text{subject to}
& & \text{Tr}\left( \mathbf{F} \left( \mathbf{M} \mathbf{M}^H + \sigma_n^2\mathbf{I}_{l} \right) { \mathbf{F} }^{ H }\right) \le {P}_{T}.  \\
\end{aligned}
\end{equation}
\begin{figure*}
\begin{align} \label{E44}
\mathbf{E}&=\left( \mathbf{V}_M\mathbf{V}_M^H + 
\frac{1}{\sigma_n^2}\mathbf{V}_M\left(\boldsymbol{\Lambda}_M-\mathbf{\Sigma}_M^H\mathbf{U}_M^H\left(\frac{\sigma_n^2}{\sigma_v^2}\mathbf{F}^H\mathbf{V}_g\mathbf{\Lambda}_g\mathbf{V}_g^H\mathbf{F}+\mathbf{I}_{q}\right) ^{-1}\mathbf{U}_M\mathbf{\Sigma}_M\right)\mathbf{V}_M^H\right )^{-1} \nonumber \\
&=\left( \mathbf{I}_m + 
\frac{1}{\sigma_n^2}\left(\boldsymbol{\Lambda}_M-\mathbf{\Sigma}_M^H\mathbf{U}_M^H\left(\frac{\sigma_n^2}{\sigma_v^2}\mathbf{F}^H\mathbf{V}_g\mathbf{\Lambda}_g\mathbf{V}_g^H\mathbf{F}+\mathbf{I}_{q}\right) ^{-1}\mathbf{U}_M\mathbf{\Sigma}_M\right)\right )^{-1} \nonumber \\
&=\left( \mathbf{I}_m + 
\frac{1}{\sigma_n^2}\mathbf{\Sigma}_M^H\left(\boldsymbol{I}_{q}-\left(\frac{\sigma_n^2}{\sigma_v^2}\mathbf{U}_M^H\mathbf{F}^H\mathbf{V}_g\mathbf{\Lambda}_g\mathbf{V}_g^H\mathbf{F}\mathbf{U}_M+\mathbf{U}_M^H\mathbf{U}_M\right) ^{-1}\right)\mathbf{\Sigma}_M\right )^{-1}.
\end{align}
\hrulefill
\end{figure*}
The above matrix-based optimization problem in terms of $\mathbf{F}$ can be converted to an equivalent scalar valued convex optimization problem via majorization theory \cite{palomar2007mimo}. Since the trace of the error covariance matrix $\mathbf{E}$ is a Schur-concave function \cite{palomar2003joint}, the MSE cost function value achieves the lower bound when $\mathrm{diag}\left(\mathbf{\widetilde{E}}\right)=\mathbf{\boldsymbol{\lambda}\left(\widetilde{E}\right)}$, and the optimal solution is obtained when the matrix is diagonalized. Let the singular value decomposition (SVD) of $\mathbf{{M}}$ be given by $\mathbf{U}_{{M}}\mathbf{\Sigma}_{{M}}\mathbf{V}_{M}^H$.
Also, the eigenvalue decomposition (EVD) of $\widetilde{\mathbf{M}}=\mathbf{{M}}^H\mathbf{{M}}$ be denoted as $\mathbf{V}_{\widetilde{M}}\mathbf{\Lambda}_{\widetilde{M}}\mathbf{V}_{\widetilde{M}}^H$, where $\mathbf{\Lambda}_{\widetilde{M}} \in \mathbb{C}^{m \times m}$ is given by $\mathbf{\Lambda}_{\widetilde{M}}=\mathrm{diag} \left[\lambda_1(\widetilde{\mathbf{M}}),\lambda_2(\widetilde{\mathbf{M}}), \cdots\lambda_m(\widetilde{\mathbf{M}})\right].$
Let the SVD of the concatenated mmWave MIMO channel $\mathbf{G}  \in \mathbb{C}^{N_R \times KN_T}$ be given as $\mathbf{{G}}=\mathbf{U}_g\mathbf{\Sigma}_g\mathbf{V}_g^H.$
Furthermore, the EVD of $\mathbf{G}^H\mathbf{G}$ may be derived as $ \widetilde{\mathbf{G}}=
\mathbf{G}^H\mathbf{G}=\mathbf{V}_g\mathbf{\Lambda}_g\mathbf{V}_g^H$,
where $\mathbf{\Lambda}_g \in \mathbb{C}^{KN_T \times KN_T}$ is given by $\mathbf{\Lambda}_g=\mathbf{\Sigma}^H_g\mathbf{\Sigma}_g=\mathrm{diag}\left[\sigma^2_1\left({\mathbf{{G}}}\right),\sigma^2_2\left({\mathbf{{G}}}\right), \cdots, \sigma^2_{KN_T}\left({\mathbf{{G}}}\right)\right],$
 with the first $L$ non-zero eigenvalues arranged in decreasing order and $\sigma_l\left(\mathbf{G}\right)$ denotes the $l$th singular value. The above decompositions can be substituted into the expression of the error covariance matrix $\mathbf{E}$ in \eqref{err_cov_noisy}, which allows the final expression to be simplified to the one shown in \eqref{E44}. From Appendix \ref{maj}, in order to diagonalize \eqref{E44}, one can choose the following structure
\begin{equation} \label{E48}
\mathbf{F}\mathbf{U}_M=\mathbf{V}^1_g\mathbf{\Sigma},
\end{equation}
where we have a diagonal matrix $\mathbf{\Sigma} \in \mathbb{R}^{m \times m}$  and $\mathbf{V}^1_g$ is comprised of the $m$ dominant left singular vectors of $\mathbf{V}_g$.
The error covariance matrix from \eqref{E44} is now diagonalized by substituting \eqref{E48} into \eqref{E44}, and the MSE is given in \eqref{E49}, where we have:
\begin{align*}
   \mathbf{V}_g^H\mathbf{V}^1_g= \begin{bmatrix}
   \mathbf{I}_{m}\\
\mathbf{0}_{(KN_T-m) \times q}
\end{bmatrix},\mathbf{\widetilde{\Lambda}}_g=\mathbf{{\Lambda}}_g\left(1:m,1:m\right). 
\end{align*}
\begin{figure*}
\begin{align} \label{E49}
\text{MSE}&=\text{Tr}\left(\left( \mathbf{I}_m + 
\frac{1}{\sigma_n^2}\mathbf{\Sigma}_M^H\left(\boldsymbol{I}_{q}-\left(\frac{\sigma_n^2}{\sigma_v^2}\mathbf{\Sigma}^H{\mathbf{V}^1_g}^H\mathbf{V}_g\mathbf{\Lambda}_g\mathbf{V}_g^H\mathbf{V}^1_g\mathbf{\Sigma}+\mathbf{I}_{q}\right) ^{-1}\right)\mathbf{\Sigma}_M\right )^{-1} \right)\nonumber \\
&=\text{Tr}\left(\left( \mathbf{I}_m + 
\frac{1}{\sigma_n^2}\mathbf{\Sigma}_M^H\left(\boldsymbol{I}_{q}-\left(\frac{\sigma_n^2}{\sigma_v^2}\mathbf{\Sigma}^H\mathbf{\widetilde{\Lambda}}_g\mathbf{\Sigma}+\mathbf{I}_{q}\right) ^{-1}\right)\mathbf{\Sigma}_M\right )^{-1}\right) \nonumber \\
&=\text{Tr}\left(\left( \mathbf{I}_m + 
\frac{1}{\sigma_n^2}\mathbf{\widetilde{\Sigma}}_M^H\left(\boldsymbol{I}_{m}-\left(\frac{\sigma_n^2}{\sigma_v^2}\mathbf{\widetilde{\Sigma}}^H\mathbf{\widetilde{\Lambda}}_g\mathbf{\widetilde{\Sigma}}+\mathbf{I}_{m}\right) ^{-1}\right)\mathbf{\widetilde{\Sigma}}_M\right )^{-1}\right).
\end{align}
\hrulefill
\end{figure*}
The matrix $\mathbf{\Sigma} \in \mathbb{R}^{m \times m}$  is given as $\mathbf{\Sigma}= 
  \left[ \mathrm{diag}\left( \boldsymbol{\mathrm{p}}\right) \right]^{\frac{1}{2}}_{m \times m}$
where ${\mathbf{p}}=\left[{p_1},{p_2},...,{p}_m\right]^T $ is the power allocation vector. Thus, the MSE can now be written as
\begin{equation} \label{E50}
\text{MSE}= \sum_{l=1}^m \frac{\sigma_v^2+\sigma_n^2p_l\sigma^2_l(\widetilde{\mathbf{G}})}{\sigma_v^2+\left(\sigma_n^2+\lambda_l(\widetilde{\mathbf{M}})\right)p_l\sigma^2_l(\widetilde{\mathbf{G}})}.
\end{equation}
Substituting the TPC matrix $\mathbf{F}$ and using EVD of $\widetilde{\mathbf{M}}$, the total transmit power can be expressed as
\begin{align} \label{E51}
&\text{Tr}\left( \mathbf{F} \left( \mathbf{M} \mathbf{M}^H + \sigma_n^2\mathbf{I}_{q} \right) { \mathbf{F} }^{ H }\right) \nonumber \\  
&=\text{Tr}\left( \mathbf{V}^1_g\mathbf{\Sigma} \left(\mathbf{\Lambda}_{\widetilde{M}}^H + \sigma_n^2\mathbf{I}_{q} \right)\mathbf{\Sigma}^H (\mathbf{V}^1_g)^H\right) \nonumber \\
&= \sum_{l=1}^mp_l\left(\lambda_l(\widetilde{\mathbf{M}}) + \sigma_n^2\right).
\end{align}
Using \eqref{E50} and \eqref{E51}, the optimization problem of minimizing the MSE of the estimate of the parameter vector $\boldsymbol{\theta}$ subject to a power constraint can be equivalently formulated as
\begin{align}
 \underset{\mathbf{p}}{\text{minimize}}
\  &\sum_{l=1}^m \frac{\sigma_v^2+\sigma_n^2p_l\sigma^2_l(\widetilde{\mathbf{G}})}{\sigma_v^2+\left(\sigma_n^2+\lambda_l(\widetilde{\mathbf{M}})\right)p_l\sigma^2_l(\widetilde{\mathbf{G}})} \nonumber\\
\text{subject to}
\   &\sum\limits_{l=1}^m p_l\left(\lambda_l(\widetilde{\mathbf{M}}) + \sigma_n^2\right) \le P_T, \nonumber\\
& p_l \geq 0, 1 \leq l \leq m.
    \end{align}
 The above problem is convex in terms of the power allocation vector $\mathbf{p}$. Employing the Karush-Kuhn-Tucker (KKT) conditions \cite{boyd2004convex}, the optimal value of $p_l$ is obtained as
   \begin{equation}
   p_l=\frac{\left(\mu\sqrt{\frac{\sigma_v^2\lambda_l(\widetilde{\mathbf{M}})\sigma^2_l(\widetilde{\mathbf{G}})}{\left(\sigma_n^2+\lambda_l(\widetilde{\mathbf{M}})\right)}}-\sigma_v^2\right)^{+}}{\left(\sigma_n^2+\lambda_l(\widetilde{\mathbf{M}})\right)\sigma^2_l(\widetilde{\mathbf{G}})},
   \end{equation}
where the Lagrangian multiplier $\mu$ is given by
   \begin{equation}
   \mu=\frac{P_T+\sum_{l=1}^m\frac{\sigma_v^2}{\sigma^2_l(\widetilde{\mathbf{G}})}}{\sum_{l=1}^m\sqrt{\frac{\sigma_v^2\lambda_l(\widetilde{\mathbf{M}})}{\left(\sigma_n^2+\lambda_l(\widetilde{\mathbf{M}})\right)\sigma^2_l(\widetilde{\mathbf{G}})}}}.
   \end{equation}
The matrix $ \mathbf{{F}}\mathbf{U}_M$ is obtained by substituting the optimal values $p_l$ into the expression of $\mathbf{\Sigma}$ in \eqref{E48}. 
The individual fully-digital TPCs $\mathbf{F}_k$, $1 \leq k \leq K$, can be determined as
\begin{equation} \label{E55}
\mathbf{F}_k=\mathbf{V}_{g,k}^1\boldsymbol{\Sigma}\left(\mathbf{U}_{M,k}\right)^{\dagger},
\end{equation}
where $\mathbf{V}^1_{g,k}$ denotes the sub-matrix of $\mathbf{V}^1_{g}$ comprised of the rows from $(k-1)N_T + 1$ to $kN_T$, whereas $\mathbf{U}_{M,k}$ is the sub-matrix of $\mathbf{U}_{M}$ comprising the rows from $\sum_{j=1}^{k-1}q_j + 1$ to $\sum_{j=1}^{k}q_j$. 

The RF TPC $\mathbf{F}_{\text{RF},k}$ and baseband TPC $\mathbf{F}_{\text{BB},k}$ can now be designed from the optimal fully-digital precoder $\mathbf{F}_k$ of the $k$th sensor as follows
\begin{equation}
\begin{aligned} \label{E34}
\underset{\mathbf{F}_{\text{RF},k},{\mathbf{F}_{\text{BB},k}}}{\text{minimize}} \quad &\left\Vert{ \left(\mathbf{F}_k-\mathbf{F}_{\text{RF},k}{\mathbf{F}}_{\text{BB},k}\right)}\right\Vert_{F}^2 \\
\textrm{subject to}
\quad &  \vert \mathbf{F}_{\text{RF},k}(s,t)\vert = \frac{1}{\sqrt{N_T}},\\
\end{aligned}
\end{equation}
where $\left\Vert \mathbf{Z}\right\Vert_F$ denotes the Frobenius norm \cite{el2014spatially} of the matrix $\mathbf{Z}$. Due to the constant magnitude problem described earlier, the problem is non-convex and intractable. However, we can now implement the ensuing remarks which allows us to simplify the hybrid TPC design. Please note that our system model is different from that of cellular mmWave MIMO systems that employ either point-to-point MIMO or multi-user MIMO communication systems. In our work, we consider a coherent MAC channel where the signals gleaned from the sensor nodes are superimposed at the FC. Therefore, the detailed proofs that demonstrate the fact that the fully-digital TPC lies in the row space of the corresponding MIMO channel is also an important novel contribution of this work. From \eqref{E10}, it can be inferred that the row and column spaces  of the mmWave MIMO channel $\mathbf{G}_k$ between the $k$th sensor and FC constitute a subset of the column spaces of the transmit and receive array response matrices $\mathbf{A}_{T,k}$ and $\mathbf{A}_R$, respectively, i.e.
\begin{equation} \label{E36}
\mathcal{R}\left(\mathbf{G}_k\right) \subseteq \mathcal{C}\left(\mathbf{A}_{T,k}\right)\ \  \text{and} \ \ \mathcal{C}\left(\mathbf{G}_k\right) \subseteq \mathcal{C}\left(\mathbf{A}_{R}\right).
\end{equation}
Furthermore, interestingly, the relationship between the column-space of the fully-digital TPC $\mathbf{F}_k$ at the $k$th sensor and the column-space of the transmit array response matrix $\mathbf{A}_{T,k}$ is presented in the theorem below.
\begin{theorem} \label{Proof_subspace}
The column-space of the optimal fully-digital TPC $\mathbf{F}_k$ at the $k$th sensor lies in the column-space of the transmit array response matrix $\mathcal{C}\left(\mathbf{A}_{T,k}\right)$, i.e.,
\begin{equation} 
\mathcal{C}\left(\mathbf{F}_k\right) \subseteq \mathcal{C}\left(\mathbf{A}_{T,k}\right).
\end{equation}
\end{theorem}
\begin{proof}
The proof is given in Appendix \ref{TPCproof}.
\end{proof}
Since there are only $N_{\text{RF}}^s$ RF chains, the RF TPC $\mathbf{F}_{\text{RF},k}$ of size $N_T \times N_{\text{RF}}^s$ corresponding to the $k$th sensor can be obtained by choosing $N_{\text{RF}}^s$ columns of the transmit array response matrix $\mathbf{A}_{T,k} \in \mathbb{C}^{N_T \times L}$, since the elements of $\mathbf{A}_{T,k}$ satisfy the constant magnitude constraint.
The pertinent optimization problem in \eqref{E34} can thus be reformulated as
\begin{equation}
\begin{aligned} \label{E40}
 \mathop{\text{minimize}}_{\widetilde{\mathbf{F}}_{\text{BB},k}}  \quad & \left\Vert{ \left(\mathbf{F}_k-\mathbf{A}_{T,k}\widetilde{\mathbf{F}}_{\text{BB},k}\right)}\right\Vert_{F}^2 \\
 \textrm{subject to}  \quad & 
\left\Vert \mathrm{diag}( \widetilde{\mathbf{F}}_{\text{BB},k}\widetilde{\mathbf{F}}_{\text{BB},k}^H) \right\Vert_0=N_{\text{RF}}^s ,
\end{aligned}
\end{equation}
where $\widetilde{\mathbf{F}}_{\text{BB},k} \in \mathbb{C}^{L \times q_k}$ is the intermediate baseband TPC corresponding to $\mathbf{A}_{T,k}$ and $\left\Vert\mathbf{Z}\right\Vert_0$ denotes the $l_0$ norm \cite{el2014spatially} of the matrix $\mathbf{Z}$. Accordingly, $\widetilde{\mathbf{F}}_{\text{BB},k}$ is block sparse in nature where $N_{\text{RF}}^s$ out of $L$ rows are non-zero. The baseband TPC $\mathbf{F}_{\text{BB},k}$ corresponds to the $N_{\text{RF}}^s$ non-zero rows of $\widetilde{\mathbf{F}}_{\text{BB},k}$. Therefore, the resultant MMV-based sparse signal recovery problem for our
TPC design can then be solved using the SOMP technique described in Algorithm \ref{alg:algorithm1}, where the matrices $\mathbf{A}_{T,k}$ and $\widetilde{\mathbf{F}}_{\text{BB},k}$ act as auxiliary
variables for the design of the RF and BB TPCs $\mathbf{F}_{\text{RF},k}$ and $\mathbf{F}_{\text{BB},k}$, respectively.
\begin{algorithm}[ht]
\caption{ Simultaneous orthogonal matching pursuit (SOMP) algorithm for the transceiver design}\label{alg:algorithm1}
\begin{algorithmic}[1]
\Require{$\left\lbrace \mathbf{F}_k\right\rbrace \forall k$ and $N_{\text{RF}}^s$ }
\For{$1 \leq k \leq K$}
\State $\widehat{\mathbf{F}}_{\text{RF},k}=[\  ]$
\State $\mathbf{F}_{\text{res}}=\mathbf{F}_k$
\For{$c\leq N_{\text{RF}}^{s}$}
   \State$\boldsymbol{\Psi}=\mathbf{A}_{T,k}^H\mathbf{F}_{\text{res}}$
   \State $l={\textrm{arg max}}_{n=1, ..., L}(\boldsymbol{\Psi}\boldsymbol{\Psi}^H)_{n,n}$ 
   \State $\widehat{\mathbf{F}}_{\text{RF},k}=[\widehat{\mathbf{F}}_{\text{RF},k}\mid \mathbf{A}_{T,k}^{(l)}]$
   \State $\widehat{\mathbf{F}}_{\text{BB},k}=(\widehat{\mathbf{F}}_{\text{RF},k}^H\widehat{\mathbf{F}}_{\text{RF},k})^{-1}\widehat{\mathbf{F}}_{\text{RF},k}^H\mathbf{F}_k$
   \State $\mathbf{F}_{\text{res}}=\frac{\mathbf{F}_k-\widehat{\mathbf{F}}_{\text{RF},k}\widehat{\mathbf{F}}_{\text{BB},k}}{\|\mathbf{F}_k-\widehat{\mathbf{F}}_{\text{RF},k}\widehat{\mathbf{F}}_{\text{BB},k}\|_F}$
\EndFor
\State $\mathbf{F}_{\text{RF},k}=\widehat{\mathbf{F}}_{\text{RF},k}$
\State $\mathbf{F}_{\text{BB},k}=\widehat{\mathbf{F}}_{\text{BB},k}$
\State $\mathbf{F}_{\text{RF}}=\text{blkdiag}\left(\mathbf{F}_{\text{RF}},\mathbf{F}_{\text{RF},k}\right)$
\State $\widehat{\mathbf{F}}_{\text{BB}}=\text{blkdiag}\left(\widehat{\mathbf{F}}_{\text{BB}},\mathbf{F}_{\text{BB},k}\right)$
\EndFor
\State $\mathbf{F}_{\text{BB}}=\sqrt{P_T}\frac{\widehat{\mathbf{F}}_{\text{BB}}}{\vert\vert\mathbf{F}_{\text{RF}}\widehat{\mathbf{F}}_{\text{BB}}\left(\mathbf{M}\mathbf{M}^H + \sigma_n^2\mathbf{I}_{q}\right)^{\frac{1}{2}}\vert\vert_F}$
\end{algorithmic}
\end{algorithm}
For the $k$th sensor, the algorithm requires as inputs the fully-digital TPC $\mathbf{F}_k$ determined in \eqref{E33} and the number of RF chains $N_{\text{RF}}^s$. Algorithm \ref{alg:algorithm1} starts with Step (5) by evaluating the projection of each column of $\mathbf{A}_{T,k}$ on every column of the residual matrix $\mathbf{F}_{\text{res}}$. In Step (6), one chooses the column of $\mathbf{A}_{T,k}$ that has the maximum $l_2$ norm of the projections on the columns of $\mathbf{F}_{\text{res}}$. In Step (7), the chosen column is appended to the RF TPC matrix $\widehat{\mathbf{F}}_{\text{RF},k}$. The baseband TPC  $\widehat{\mathbf{F}}_{\text{BB},k}$ is then computed using the least squares solution in Step (8) that best approximates the ideal fully-digital TPC $\mathbf{F}_k$. Step (9) evaluates the residual matrix $\mathbf{F}_{\text{res}}$ by subtracting the current estimate of the hybrid TPC $\widehat{\mathbf{F}}_{\text{RF},k}\widehat{\mathbf{F}}_{\text{BB},k}$ from the ideal TPC $\mathbf{F}_k$, followed by normalization. After $N_{\text{RF}}^s$ iterations, one obtains the final baseband and RF TPC $\mathbf{F}_{\text{BB},k},\mathbf{F}_{\text{RF},k}$ at the $k$th sensor. Step (13) and (14) then return the block diagonal baseband TPC $\widehat{\mathbf{F}}_{\text{BB}}$ and RF TPC $\mathbf{F}_{\text{RF}}$. ${\mathbf{F}}_{\text{BB}}$ is obtained in Step (16) after ensuring that the total transmit power constraint in \eqref{eq5} is satisfied. This completes the hybrid TPC design procedure for the mmWave MIMO WSN. The following subsection describes hybrid TPC design incorporating individual sensor power constraints.

\subsection{TPC design subject to individual sensor power constraints} \label{SectionIVB}
The transmit power constraint at every sensor $k$ is given by \eqref{eq9} as 
\begin{equation}\label{59}
\text{Tr}\left( \mathbf{F}_k \left( \mathbf{M}_k \mathbf{M}_k^H + \sigma_n^2\mathbf{I}_{q_k} \right) { \mathbf{F} }^{ H }_k\right) \le P_k,
\end{equation}
where $\mathbf{F}_k$ is the fully-digital TPC. 
Similar to the TPC design problem formulated in \eqref{ee1} in the previous subsection for a total power budget, the corresponding problem for this scenario associated with each sensor constrained by $P_k$ can be developed as
\begin{equation}
\begin{aligned}
& \underset{\mathbf{{F}}}{\text{minimize}}
& &   \text{Tr}\left( \mathbf{E}\right)    \\
&\text{subject to}
& & \text{Tr}\left( \mathbf{F}_k \left( \mathbf{M}_k  \mathbf{M}_k^H + \sigma_n^2\mathbf{I}_{q_k} \right) { \mathbf{F} }^{ H }_k\right) \le P_k, \ \forall k.\\
\end{aligned}
\end{equation}
As shown in Appendix \ref{maj}, since the trace of the error covariance matrix $\mathbf{E}$ is a Schur-concave function, the MSE achieves the lower bound when the matrix $\mathbf{E}$ is diagonalized. Thus, the fully-digital TPC structure follows the structure of \eqref{E48}, which can be subsequently written as
\begin{align} \label{TPC_ind_noisy}
&\mathbf{F}\mathbf{U}_M=\mathbf{V}^1_g\mathbf{\Sigma} \nonumber \\
&\Rightarrow  \begin{bmatrix}
\mathbf{F}_1 & \mathbf{0}& \cdots&\mathbf{0}\\
\mathbf{0}& \mathbf{F}_2 &\cdots &\mathbf{0} \\
\vdots& & \ddots &\vdots \\
\mathbf{0}&\mathbf{0} & & \mathbf{F}_K
\end{bmatrix}\begin{bmatrix}
\mathbf{U}_{M,1} \\
\mathbf{U}_{M,2}\\
\vdots\\
\mathbf{U}_{M,K}
\end{bmatrix} =\begin{bmatrix}
\mathbf{V}_{g,1}^1 \\
\mathbf{V}_{g,2}^1\\
\vdots\\
\mathbf{V}_{g,K}^1\end{bmatrix}\mathbf{\Sigma} \\ \nonumber
&\Rightarrow \mathbf{F}_k\mathbf{U}_{M,k}=\mathbf{V}_{g,k}^1\mathbf{\Sigma} \ \ \forall k.
\end{align} 
From the SVD of $\mathbf{M}$, the measurement matrix $\mathbf{M}_k$ of the $k$th sensor can be expressed as
\begin{equation}
\mathbf{M}_k=\mathbf{U}_{M,k}\boldsymbol{\Sigma}_M\mathbf{V}_{M}^H.
\end{equation}
Considering the fact that $\mathbf{U}_{M,k} \mathbf{U}_{M,k}^H=\mathbf{I}_{q_k}$, the per sensor transmit power in Eq. \eqref{59} can be written in the scalar form of:
\begin{align} \label{E68}
&\text{Tr}\left( \mathbf{F}_k \left( \mathbf{U}_{M,k}\boldsymbol{\Sigma}_M\mathbf{V}_{M}^H\mathbf{V}_{M}\boldsymbol{\Sigma}^H_M\mathbf{U}_{M,k}^H  + \sigma_n^2\mathbf{I}_{q_k} \right) { \mathbf{F} }^{ H }_k\right)  \nonumber \\
&=\text{Tr}\left( \mathbf{F}_k  \mathbf{U}_{M,k}\left(\boldsymbol{\Sigma}_M\boldsymbol{\Sigma}^H_M + \sigma_n^2 \mathbf{I}_{q}\right)\mathbf{U}_{M,k}^H  { \mathbf{F} }^{ H }_k\right)  \nonumber \\
&\stackrel{\text{a}}{=}\text{Tr}\left(\underbrace{(\mathbf{V}_{g,k}^1)^{ H } \mathbf{V}_
{g,k}^1}_{\boldsymbol{\Phi}_k}  \boldsymbol{\Sigma}\left(\boldsymbol{\Sigma}_M\boldsymbol{\Sigma}_M^H + \sigma_n^2 \mathbf{I}_{q}\right)\boldsymbol{\Sigma}^H\right) \nonumber \\ 
&=\sum_{l=1}^mp_l\left(\lambda_l(\widetilde{\mathbf{M}}) + \sigma_n^2\right)\left[\boldsymbol{\Phi}_k\right]_{ll},
\end{align}
where (a) follows from \eqref{TPC_ind_noisy}. Hence, the scalar-valued optimization problem of minimizing
the MSE of the estimate of the parameter vector $\boldsymbol{\theta}$ can be
equivalently formulated using the objective in  \eqref{E50} and constraint in as,
\begin{align}
 \underset{\mathbf{p}}{\text{minimize}}
\  &\sum_{l=1}^m\frac{1}{\sigma_n^2+\lambda_l(\widetilde{\mathbf{M}})} \times \\ \nonumber
&\left(\sigma_n^2+ \frac{\sigma_v^2\lambda_l(\widetilde{\mathbf{M}})}{\sigma_v^2+\left(\sigma_n^2+\lambda_l(\widetilde{\mathbf{M}})\right)p_l\sigma^2_l(\widetilde{\mathbf{G}})}\right) \nonumber \\
\text{subject to}
\   &\sum_{l=1}^mp_l\left(\lambda_l(\widetilde{\mathbf{M}}) + \sigma_n^2\right)\left[\boldsymbol{\Phi}_k\right]_{ll} \le P_k,
  \  \forall k, \nonumber\\
& p_l \geq 0, 1 \leq l \leq m.
\end{align}
The optimal $\mathbf{p}$ can be readily determined using a convex solver. Finally, the fully-digital optimal TPC at the $k$th sensor is formulated in \eqref{E55}. From Theorem \ref{Proof_subspace} and Appendix \ref{TPCproof}, one observes that $\mathcal{C}\left(\mathbf{F}_k\right) \subseteq \mathcal{C}\left(\mathbf{A}_{T,k}\right) $. Thus, the RF TPC $\mathbb{F}_{\text{RF},k}$ at the $k$th sensor can be designed by choosing the columns of the transmit array response matrix $\mathbf{A}_{T,k}$.
 The corresponding baseband and RF components from the fully-digital hybrid TPC are determined by again solving the problem in \eqref{E40}.
Finally, Algorithm \ref{alg:algorithm1} can once again be employed for the design of the RF and baseband TPCs $\mathbf{F}_{\text{RF},k}$ and $\mathbf{F}_{\text{BB},k}$ respectively, wherein step (16) is omitted and the computation $\mathbf{F}_{\text{BB},k}=\sqrt{P_k}\frac{\widehat{\mathbf{F}}_{\text{BB},k}}{\vert\vert\mathbf{F}_{\text{RF},k}\widehat{\mathbf{F}}_{\text{BB},k}\left(\mathbf{M}_k\mathbf{M}_k^H + \sigma_n^2\mathbf{I}_{q_k}\right)^{\frac{1}{2}}\vert\vert_F}$ is performed at step (12) to meet the individual power constraints at the sensors. The TPC design of the scenario associated with high-SNR sensor observations, i.e. $\text{SNR}_n \gg 1$,
is presented in Appendix \ref{noiseless}. The algorithm of our hybrid RC design at the FC is presented next. 
\subsection{Hybrid RC design}
After designing the hybrid TPCs $\left\lbrace\mathbf{F}_{\text{RF},k},\mathbf{F}_{\text{BB},k}\right\rbrace_{k=1}^K$, one
can design hybrid MMSE RCs for the FC using the optimization
problem formulated below. Note that the combiner design
is common for both per-sensor and total power constraint
problems: 
\begin{equation} \label{eq182}
\begin{aligned}
\underset{\mathbf{W}_{\text{RF}},\mathbf{W}_{\text{BB}}}{\text{minimize}} \quad &  \mathbb{E}\left \{ \left\Vert \boldsymbol{\theta}-\mathbf{W}_{\text{BB}}^{H}\mathbf{W}_{\text{RF}}^{H}\mathbf{y} \right\Vert_2^2\right \}  \\
\textrm{subject to} \quad & \vert \mathbf{W}_{\text{RF}}(s,t)\vert = \frac{1}{\sqrt{N_R}}.
\end{aligned}
\end{equation}
Note that, the solution of the unconstrained version of \eqref{eq182} is the LMMSE combiner as given in \eqref{eq44}. Furthermore, using the mathematical manipulations detailed in Appendix \ref{RCproof}, the problem \eqref{eq182} is seen to be equivalent to 
\begin{equation} \label{eq40}
\begin{aligned}
\underset{\mathbf{W}_{\text{RF}},\mathbf{W}_{\text{BB}}}{\text{minimize}}\quad & \mathbb{E}\left \{\left\Vert \mathbf{R}_{yy}^{\frac{1}{2}}\left(\mathbf{W}-\mathbf{W}_{\text{RF}}\mathbf{W}_{\text{BB}}\right)\right\Vert_{F}^2\right \}  \\
 \textrm{subject to} \quad & \vert \mathbf{W}_{\text{RF}}(s,t)\vert = \frac{1}{\sqrt{N_R}},
\end{aligned}
\end{equation}
where $\mathbf{R}_{yy} = \mathbb{E}\left\{ \mathbf{y}\mathbf{y}^{H}\right\} \in \mathbb{C}^{N_R \times N_R}$ is the covariance matrix of the signal $\mathbf{y}$ received by the FC. The hybrid RC comprised of the RF and baseband RCs $\mathbf{W}_{\text{RF}}$ and $\mathbf{W}_{\text{BB}}$, respectively, can once again be designed from the ideal LMMSE RC $\mathbf{W}$ in \eqref{eq44} as the solution to the optimization problem in \eqref{eq40}. The pertinent covariance matrix $\mathbf{R}_{yy}\in \mathbb{C}^{N_R \times N_R}$ of the received signal $\mathbf{y}$ for this scenario is given by 
\begin{align}
\mathbf{R}_{yy}= {\mathbf{G}}\mathbf{F}\left(\mathbf{M}{\mathbf{M}}^H+\mathbf{R}_n\right)\mathbf{F}^H\mathbf{G}^H + \sigma_v^2\mathbf{I}_{N_R}.
\end{align}
Interestingly, one can observe from \eqref{eq44} and  \eqref{E11} that the column-space of the LMMSE combiner $\mathbf{W}$ is a subspace of the column-space of the concatenated mmWave MIMO channel $\mathbf{G}$, which in turn lies in the column-space of the receive array response matrix $\mathbf{A}_R$, i.e. $\mathcal{C}\left(\mathbf{W}\right) \subseteq \mathcal{C}\left(\mathbf{G}\right) \subseteq \mathcal{C}\left(\mathbf{A}_R\right)$. Therefore, the columns of $\mathbf{W}_{\text{RF}}$ can be chosen from the columns of the receive array response matrix $\mathbf{A}_{R}$. The pertinent optimization problem may then be recast as
\begin{align} \label{eq42}
{\mathop{\text{minimize}}_{\widetilde{\mathbf{W}}_{\text{BB}}}} \quad & \left\Vert {\mathbf{R}_{yy}^{\frac{1}{2}}\left(\mathbf{W}-\mathbf{A}_{R}\widetilde{\mathbf{W}}_{\text{BB}}\right)}\right\Vert_{F}^2 \nonumber  \\
\text{subject to} \quad & 
\  \left\Vert \mathrm{diag}( \widetilde{\mathbf{W}}_{\text{BB}}\widetilde{\mathbf{W}}_{\text{BB}}^H) \right\Vert_0=N_{\text{RF}}^{\text{FC}},  \nonumber \\
\end{align} 
where $\widetilde{\mathbf{W}}_{\text{BB}} \in \mathbb{C}^{L \times m}$ is a block sparse matrix that has $N_{\text{RF}}^{\text{FC}}$ non-zero rows. The resultant sparse signal recovery problem can once again be solved by using the SOMP procedure of Algorithm \ref{alg:algorithm1} upon replacing $\mathbf{F}_k$, $N_{\text{RF}}^s$, $\mathbf{A}_{T,k}$ by $\mathbf{R}_{yy}^{\frac{1}{2}}\mathbf{D}$, $N_{\text{RF}}^{\text{FC}}$, $\mathbf{R}_{yy}^{\frac{1}{2}}\mathbf{A}_{R}$, respectively.\par
\begin{table*}[t] 
 \caption{Comparison of computational complexity of our proposed work with \cite{behbahani2012linear}, \cite{6894624}, \cite{8332507}, \cite{liu2021hybrid}}
 \label{complexity_table}
\hspace{20pt}
 \begin{tabular}{|l|r|r|}
\hline
Algorithm & Hybrid TPCs $\left(\left\lbrace \mathbf{F}_{\text{RF},k}, \mathbf{F}_{\text{BB},k} \right\rbrace_{k=1}^K\right)$ & Hybrid RC $\left(\mathbf{W}_{\text{RF}}, \mathbf{W}_{\text{BB}}\right)$ \\
\hline
Proposed work & $\mathcal{O}\left(N_R^2N_TK\right)$ & $\mathcal{O}\bigg(K^2N_Rm^2 + \left(N_{\text{RF}}^{\text{FC}}\right)^3N_R\bigg)$ \\
\hline
BCD algorithm \cite{behbahani2012linear} &  $\mathcal{O}\bigg(N_{\text{I}}N_T^2K^2m + N_{\text{I}}N_RN_TKm + N_{\text{I}}N_T^3K\bigg)$ &$\mathcal{O}\bigg((N_{\text{RF}}^{\text{FC}})^3N_R + N_{\text{I}}N_RKm + N_{\text{I}}N_R^3\bigg)$
\\
\hline
BLUE  \cite{6894624}  & $\mathcal{O}\bigg(N_{\text{I}}\left(N_RK + K^3 \right)\bigg)$ & $\mathcal{O}\bigg(N_{\text{I}}\left(N_R^3 + (N_{\text{RF}}^{\text{FC}})^3N_R\right)\bigg)$ \\ \hline
PDD \cite{8332507} & $\mathcal{O}\bigg(N_{\text{I}}^2\left(N_{\text{RF}}^s\right)^2N_T^2K + N_{\text{I}}N_T^3K\bigg)$ & $\mathcal{O}\bigg(\left(N_{\text{RF}}^{\text{FC}}\right)^3N_RN_{\text{I}}^2\bigg)$ \\ \hline
ADMM \cite{liu2021hybrid} &$\mathcal{O}\bigg(N_{\text{I}}(N_{\text{RF}}^s)^3N_T^3K\bigg)$ & $ \mathcal{O}\bigg(KN_RN_Tm + N_{\text{I}}\left(N_{\text{RF}}^{\text{FC}}\right)^3N_R^3\bigg)$\\
\hline
\end{tabular}
\end{table*}
\subsection{Computational Complexity and Communication Overhead}
The computational complexity for calculating the hybrid TPCs $\mathbf{F}_{\text{RF},k}, \mathbf{F}_{\text{BB},k}$ corresponding to $k$th sensor is of the order of $\mathcal{O}\left(KN_TN_R^2\right)$. As it
can be observed, the complexity becomes prohibitively high as
the number of sensors in the WSN increases. The computational complexity for obtaining the hybrid RCs $\mathbf{W}_{\text{RF}}, \mathbf{W}_{\text{BB}}$ is of the order of $\mathcal{O}\left(K^2N_Rm^2 + \left(N_{\text{RF}}^{\text{FC}}\right)^3N_R\right)$. Thus, it becomes significantly high as the number of sensors $K$ and receive antennas $N_R$ increase in the network. Detailed step by-step analysis of the computational complexity is given in
our technical report \cite{TechReport}. The comparison of computational complexity of our proposed work with \cite{behbahani2012linear}, \cite{6894624}, \cite{8332507}, \cite{liu2021hybrid} is presented in Table \ref{complexity_table}.\par
It is to be noted here that the sensors are tiny battery-operated devices and have a limited battery life, and thus possess low computational capability. By contrast, the fusion center (FC) does not have such a constraint on the power, and in general will have sufficient computational and communication resources. Therefore, in our proposed scheme, the FC estimates the channels $\lbrace\mathbf{G}_k\rbrace_{k=1}^K$ using the pilot signals transmitted by the each sensor $k$ in the mmWave WSN. Subsequently, the FC designs the hybrid TPC matrices for each sensor and feeds back to each sensor its hybrid TPC. Furthermore, the FC is assumed to possess only statistical channel state information (CSI), such as the vector parameter covariance matrix $\mathbf{R}_{\theta}$, and observation noise covariance matrix $\mathbf{R}_{n}$, which can be acquired via averaging over a suitably long duration of time. Furthermore, since a quasi-static mmWave MIMO channel is considered in this work, the CSI acquired at the FC is constant over several time instants. Hence the hybrid TPCs $\mathbf{F}_{\text{BB},k}$, $\mathbf{F}_{\text{RF},k}$ does not have to be computed and fed-back very frequently. In order to further reduce the overhead, the hybrid TPC matrices $\mathbf{F}_{\text{BB},k}$ and $\mathbf{F}_{\text{RF},k}$ corresponding to the $k$ sensor can be communicated by the FC using one of the limited feedback schemes discussed in \cite{1621413}, \cite{4641946}.
\section{BCRB and centralized MMSE bound}
To benchmark the MSE performance in Section III and IV, the
BCRB and centralized MMSE bounds are determined below for the
noiseless and noisy sensor observation scenarios, respectively.
For the noiseless scenario, the BCRB \cite{van2007bayesian} for the parameter $\boldsymbol{\theta}$ is obtained as 
\begin{equation} \label{66}
\text{MSE}_{\text{BCRB}}\geq\text{Tr}\left(\left(\mathbf{I}_m + \frac{1}{\sigma_v^2}\mathbf{F}_{\text{BB}}^H\mathbf{F}_{\text{RF}}^H\mathbf{G}^H\mathbf{G}\mathbf{F}_{\text{RF}}\mathbf{F}_{\text{BB}}\right)^{-1}\right).
\end{equation}
By contrast, for the case of noisy sensor observations, the centralized MMSE benchmark represents the best achievable performance where all the sensor observations are directly available at the FC. It may be formulated as \cite{kay1993fundamentals}
\begin{equation}
\text{MSE}_{\text{cent}}=\text{Tr}\left(\left(\mathbf{I}_m + \mathbf{M}^H\mathbf{R}_n^{-1}\mathbf{M}\right)^{-1}\right).
\end{equation}
The next section presents our simulation results for characterizing the
MSE performance of the proposed hybrid TPC/RC designs conceived for mmWave MIMO WSNs. 
\section{Simulation Results}
\begin{figure*}
\centering
\subfloat[]{\includegraphics[width=0.5\linewidth]{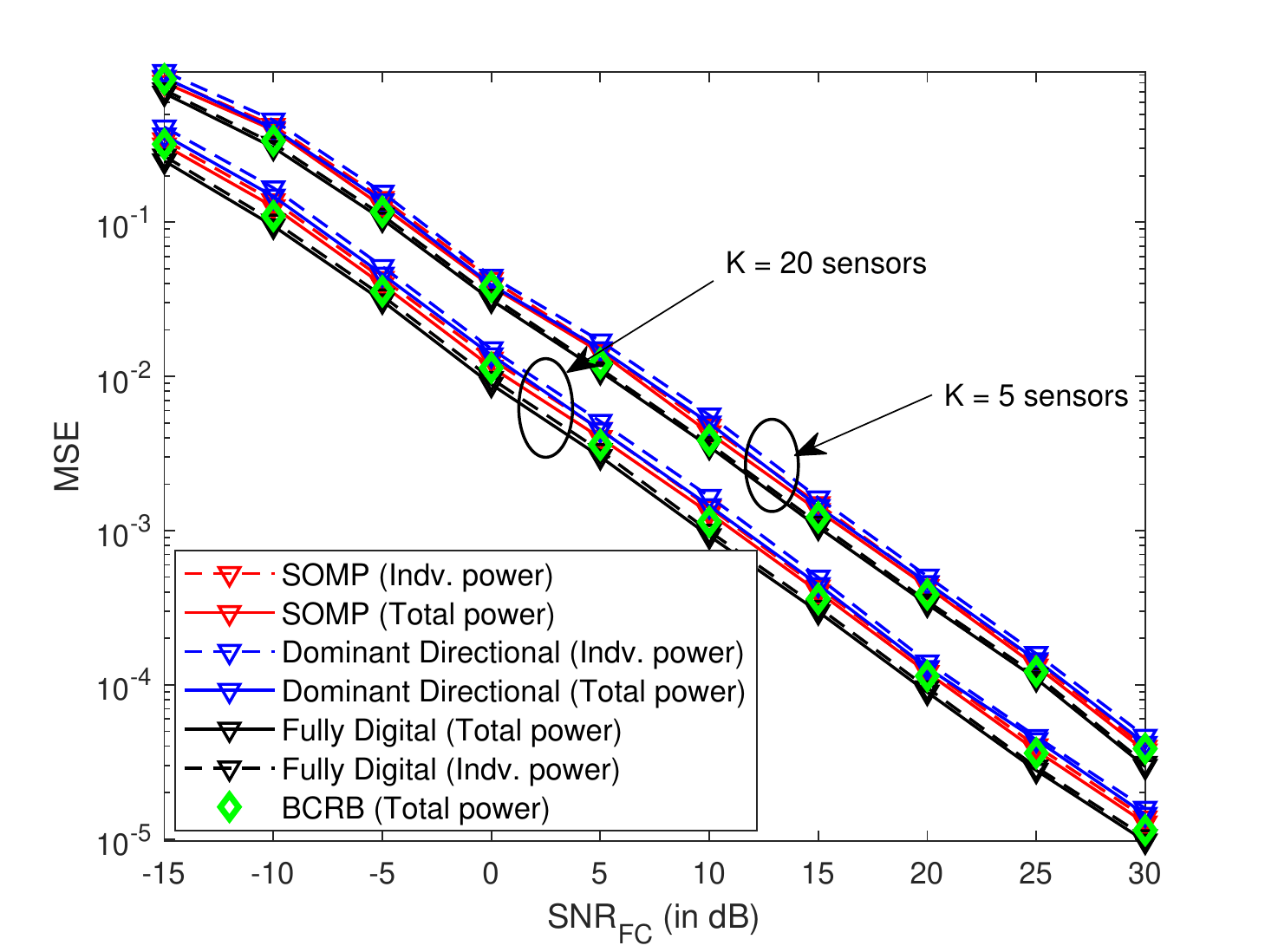}}
\hfil
\subfloat[]{\includegraphics[width=0.5\linewidth]{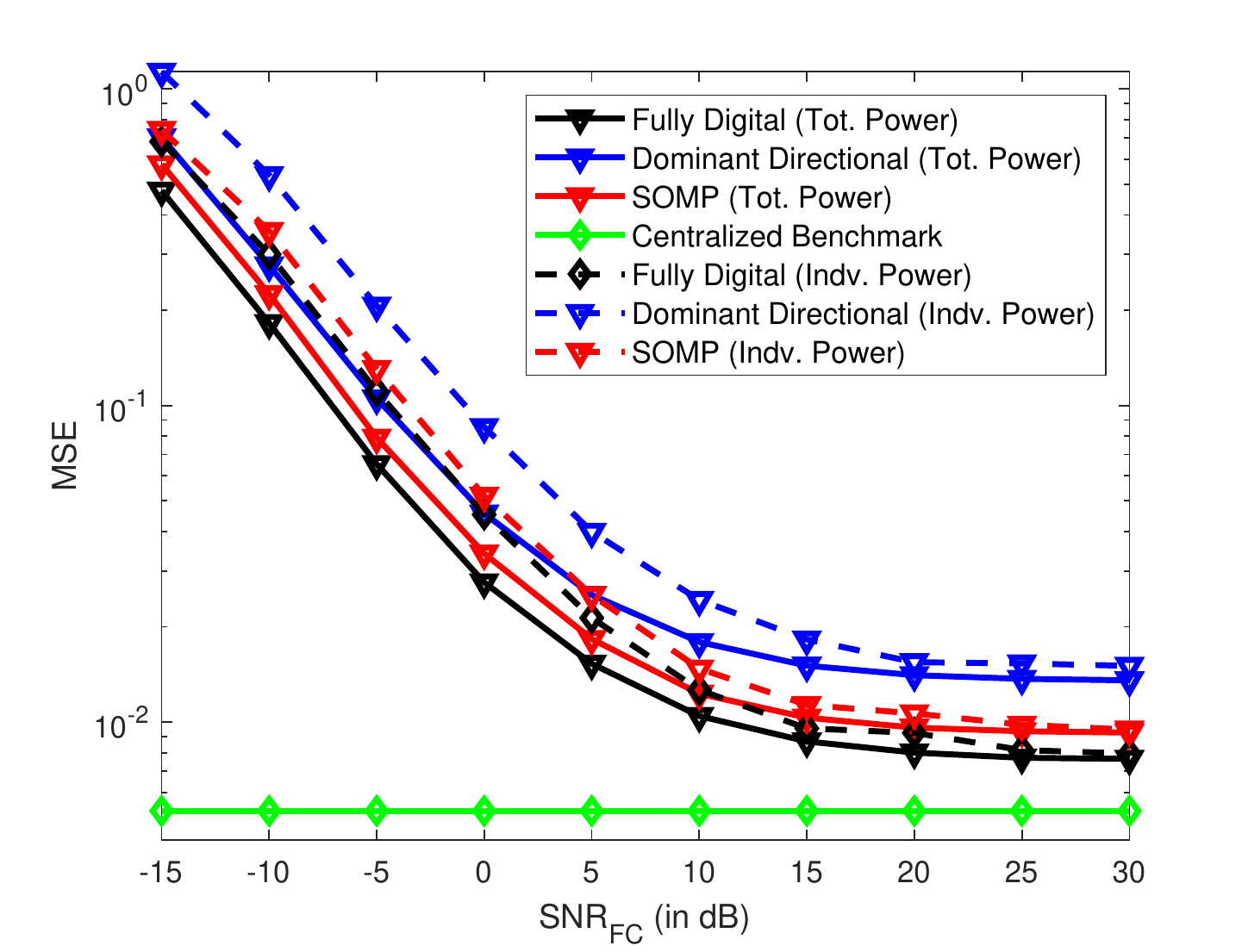}}
\caption{$\left(a\right) $ MSE versus $\text{SNR}_{\text{FC}}$ plot for noiseless sensor observations, (b) MSE versus $\text{SNR}_{\text{FC}}$ plot for noisy sensor observations}
\label{perfect_CSI}
\end{figure*}
\begin{figure*}
\centering
\subfloat[]{\includegraphics[width=0.5\linewidth]{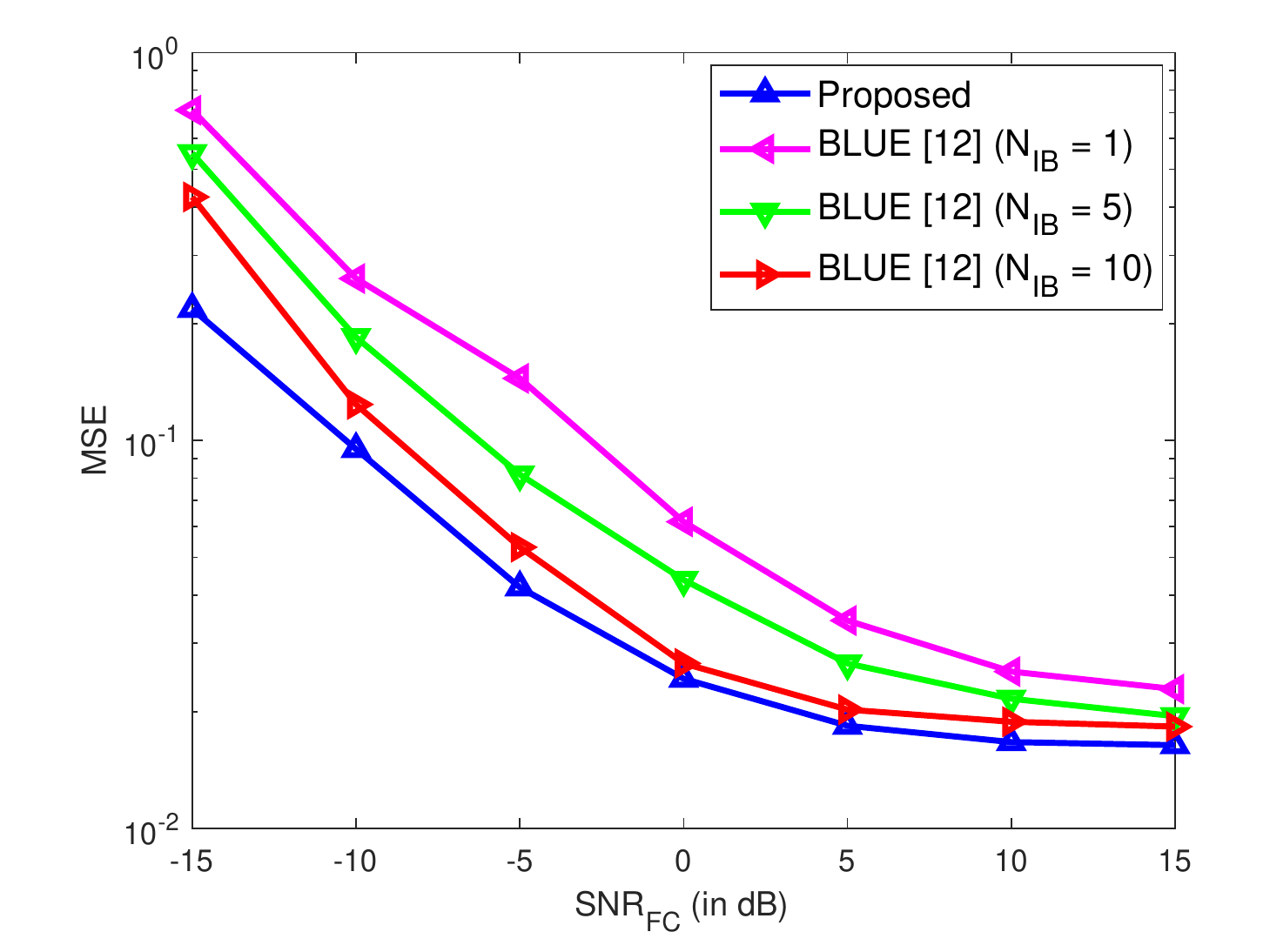}}
\hfil
\subfloat[]{\includegraphics[width=0.5\linewidth]{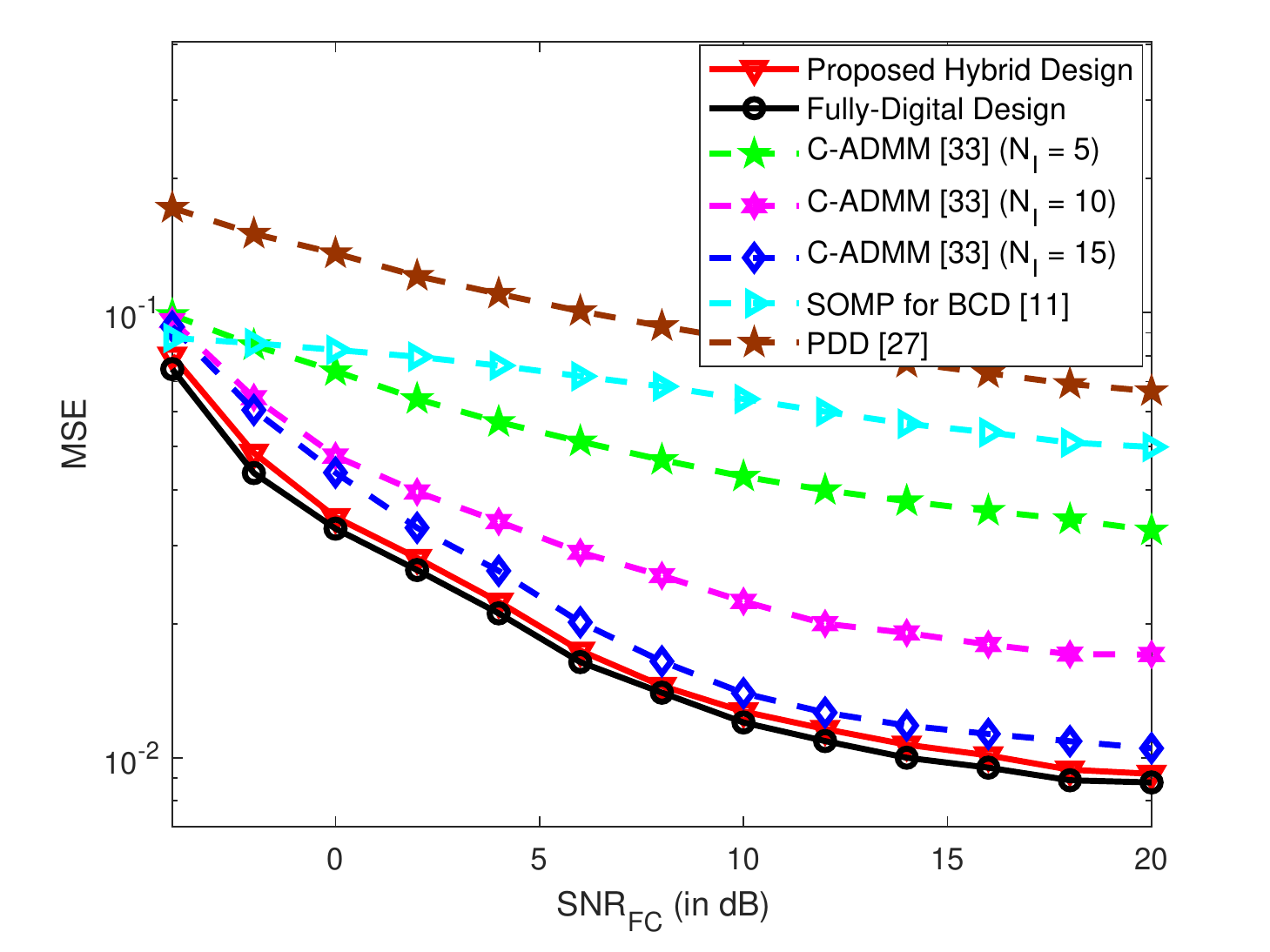}}
\caption{$\left(a\right) $ MSE versus $\text{SNR}_{\text{FC}}$ plot for scalar parameter estimation, (b) MSE versus $\text{SNR}_{\text{FC}}$ plot for vector parameter estimation}
\label{comparison}
\end{figure*}
Let us consider a mmWave MIMO WSN system comprised of $K=20$ sensors and an FC. Each sensor employs $N_T=10$ transmit antennas and $N_{\text{RF}}^s=3$ RF chains. The FC has $N_R=16$ receive antennas and $N_{\text{RF}}^{\text{FC}}=6$ RF chains. The number of multipath components is set as $L=6$. The path-gains of the mmWave MIMO channel are generated as $\mathcal{CN}(0,1)$ random variables. The AoA/AoDs are chosen randomly from the angular range $\left[0, \pi\right]$.
The number of elements in the parameter vector is set to $m=3$, while the number of measurements is set to $q_k=3$ for each of the sensors. The matrices $\mathbf{R}_v$ and $\mathbf{R}_n$ are considered to obey $\sigma_v^2\mathbf{I}_{N_R}$ and $\sigma_n^2\mathbf{I}_{q}$, respectively. The observation and FC SNRs are defined as $\text{SNR}_n=\frac{1}{\sigma_n^2}$, $\text{SNR}_{\text{FC}}=\frac{1}{\sigma_v^2}$, respectively. The $\text{SNR}_n$ is set to $10$ dB. The power budget of our WSN is set to $P_T=0\ \text{dBW}$, while the maximum transmit power of the sensors is  set to $P_k=-13\ \text{dBW}$. The MSE is plotted via Monte-Carlo simulation  by averaging over 5000 random realizations of the mmWave MIMO channel.

Fig. 3(a) shows the MSE performance of the hybrid MIMO WSN system both under total power and individual sensor power constraints for the noiseless sensor observations considered in Section III versus the $\text{SNR}_{\text{FC}}$ at the FC for various values of the number of sensors $K \in \left\lbrace 5,20\right\rbrace $. The MSE performance of the fully-digital design is also plotted to compare the efficacy of proposed hybrid transceiver design to that of the unconstrained digital design. Observe that the MSE of the proposed designs  monotonically decreases upon increasing $\text{SNR}_{\text{FC}}$ under the total power as well as individual power constraints. They are also seen to coincide with the BCRB derived in \eqref{66}. The additional
flexibility provided by the total power constraint in terms of the power allocation to each sensor results in superior
performance in comparison to individual sensor power constraints. However,
the gap in the MSE performance under both the scenarios is insignificant in the case of noiseless sensor observations for the proposed designs. This demonstrates the efficacy of the proposed design subject to individual sensor constraints, because for a given power threshold at each sensor, the MSE performance is similar to that under a total power budget for the hybrid MIMO WSN. Furthermore, the proposed designs are compared to the dominant directional TPC and RC \cite{el2014spatially}, where the parameter vector observed is steered along the dominant multipath component of the mmWave MIMO channel to yield a beamforming gain. The MSE performance of the proposed hybrid design is superior to the dominant directional precoding/combining since the proposed SOMP based design achieved the minimum MSE whereas the latter has not. Additionally, our proposed design does not require knowledge of the path gains, which is necessary for designing the dominant directional hybrid precoder/combiner. It can be observed that as the number of sensors increases, the MSE decreases. This is owing to the fact that the number of measurements  $\sum_{k=1}^Kq_k=q$ used for the estimation of the parameter $\boldsymbol{\theta}$ increases upon increasing the number of sensors,  leading to improved estimation.

Fig. 3(b) shows the MSE performance of our hybrid precoder/combiner proposed in Section IV for noisy sensor observations, versus the $\text{SNR}_{\text{FC}}$. Since the fully digital design refers to an ideal scenario, where the number of RF chains is equal to that of the antennas, it serves as a benchmark for the MSE of the hybrid transceiver design. To demonstrate the efficacy of the proposed designs, the MSE performance of the fully-digital design is plotted for both the total and individual sensor power constraints, together with the centralized MMSE benchmark. Observe that the proposed hybrid design has a performance close to that of  the fully-digital design in both the scenarios, i.e., under total and individual sensor power constraints. Moreover, with the increase of $\text{SNR}_{\text{FC}}$, the MSE is seen to decrease, approaching the centralized benchmark at high $\text{SNR}_{\text{FC}}$. The proposed designs are also compared to the dominant directional precoder/combiner under both the scenarios. Observe that the proposed SOMP based transceiver design outperforms the design based on dominant AoA/AoD selection.

%
%
%
%
%
%
%
%
%
%
%
\begin{figure*}
\centering
\subfloat[]{\includegraphics[width=0.5\linewidth]{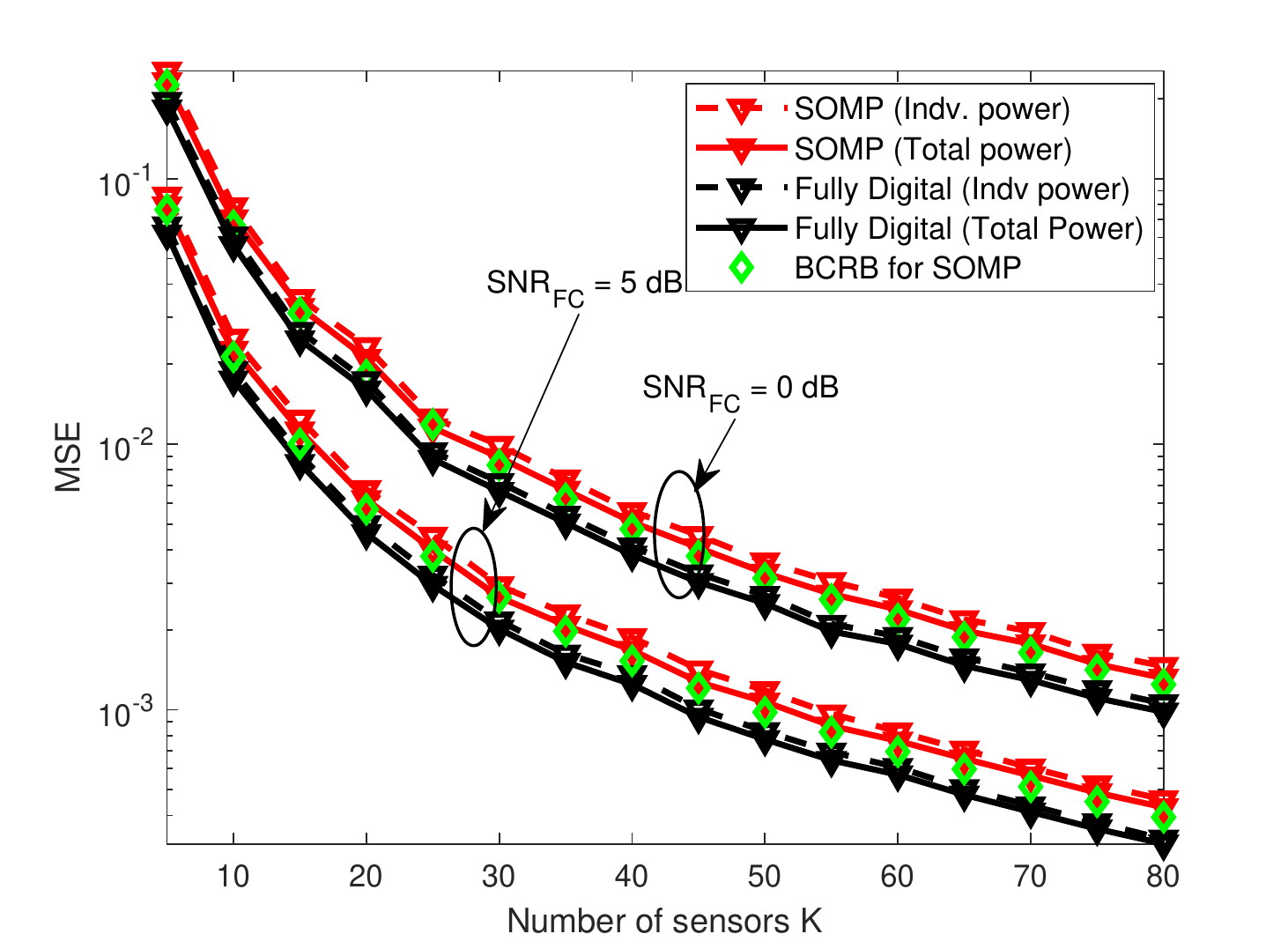}}
\hfil
\subfloat[]{\includegraphics[width=0.5\linewidth]{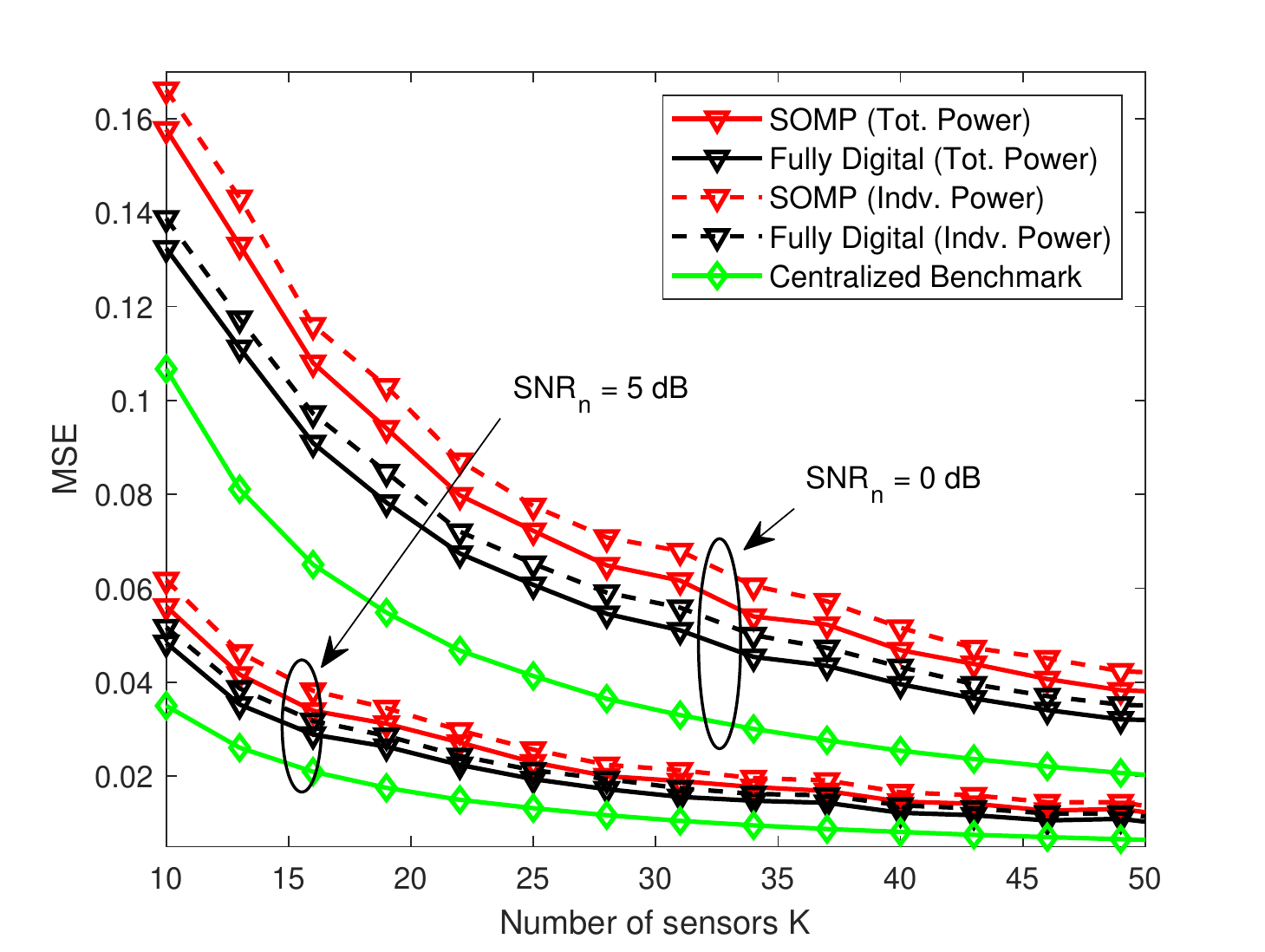}}
\caption{$\left(a\right) $ MSE versus the number of sensors $K$ for noiseless sensor observations, (b) MSE versus the number of sensors $K$ for noisy sensor observations}
\label{perfect_CSI}
\end{figure*}
Fig \ref{comparison}(a) compares the MSE performance of our hybrid TPC/ RC design of Section III with that of the SOMP-based hybrid design for the best linear unbiased estimator (BLUE) for scalar parameter estimation, i.e., $m=1$, conceived in \cite{6894624}, which designs the fully digital TPCs and the RC in an iterative fashion. The BLUE estimator is observed to result in a higher MSE in comparison to the LMMSE estimator proposed in our work because the latter scheme also incorporates the prior information of the parameter to be estimated, i.e., the covariance matrix $\mathbf{R}_{\theta}$. Observe from Table \ref{complexity_table}, that the existing RC design of \cite{6894624} is also seen to be computationally more complex than the proposed design.

Fig \ref{comparison}(b) compares the MSE performance of our hybrid TPC/ RC design of Section \ref{SectionIVB} to the hybrid TPC/ RC design in \cite{8332507}, \cite{liu2021hybrid} and to the SOMP-based hybrid design of the fully-digital TPCs/ RC matrices of \cite{behbahani2012linear} for noisy sensor observations to that of the per sensor power constraints. The figure clearly demonstrates that the MSE
performance of the PDD algorithm in \cite{8332507} is poor compared to that of the proposed design. The authors of \cite{liu2021hybrid} have proposed an ADMM-based iterative algorithm to design the hybrid TPCs/ RC. Their iterative design is observed to have a poor performance in comparison to the proposed hybrid TPC/ RC designs for a lower number of iterations and close to that of our design near $N_{\text{I}}=15$ iterations. Moreover, the proposed design has a lower computational complexity in comparison to \cite{8332507}, \cite{liu2021hybrid} as seen from Table \ref{complexity_table}.
  This demonstrates the efficacy of the proposed hybrid designs for mmWave MIMO WSNs wherein having a lower computational overhead at the FC is critical. Furthermore, the authors of \cite{behbahani2012linear} proposed an iterative algorithm for TPC/ RC designs in MIMO WSN systems. The SOMP algorithm therein is implemented to design the respective hybrid TPCs/ RC from the fully-digital TPCs/ RC matrices in \cite{behbahani2012linear} for mmWave MIMO WSNs. It can be observed that the SOMP based hybrid design also performs poorer than our proposed design for $N_{RF}^s = 3$ RF chains. Moreover, the computational complexity for \cite{behbahani2012linear}, as observed in Table \ref{complexity_table}, is  higher due to the iterative nature of the algorithm.

Fig. 5(a) depicts the MSE performance of different designs presented in Section III for noiseless sensor observations against the number of sensors $K$ in the hybrid MIMO WSN for various values of $\text{SNR}_{\text{FC}} \in \left\lbrace0, 5\right\rbrace\text{dB}$. The BCRB and MSE of the fully-digital TPC designs for both the scenarios have also been shown to benchmark the performance of the proposed designs. Upon increasing the number of sensors $K$, the MSE performance is seen to improve even at a constant total power budget of the system. This is because as the number of sensors increases, it leads to the availability of more measurements of the parameter $\boldsymbol{\theta}$, thereby increasing the estimation accuracy. The MSE performance saturates at a higher number of sensors because for a fixed total power budget of the hybrid MIMO WSN, having more measurements will no longer lead to a remarkable performance improvement.

Fig. 5(b) shows the MSE performance of the different designs presented in Section IV for noisy sensor observations against the number of sensors $K$ in the hybrid MIMO WSN for different values of $\text{SNR}_n \in \left\lbrace0, 5\right\rbrace \text{dB}$ at each noisy sensor at $\text{SNR}_{\text{FC}}=20$ dB. The MSE performance improves upon increasing the number of sensors $K$,  even at a constant power budget of the system, which follows a trend similar to the previous scenario subject to noiseless sensor observations. It is to be noted that due to the observation noise at each sensor, the gap between the centralized benchmark and the fully-digital design becomes higher compared to the scenario having noiseless sensor observations. Upon reducing the observation noise power $\sigma_n^2$, the MSE performance improves, consequently reducing the gap with respect to the centralized benchmark.

\begin{figure} 
\centering
\includegraphics[width=\linewidth]{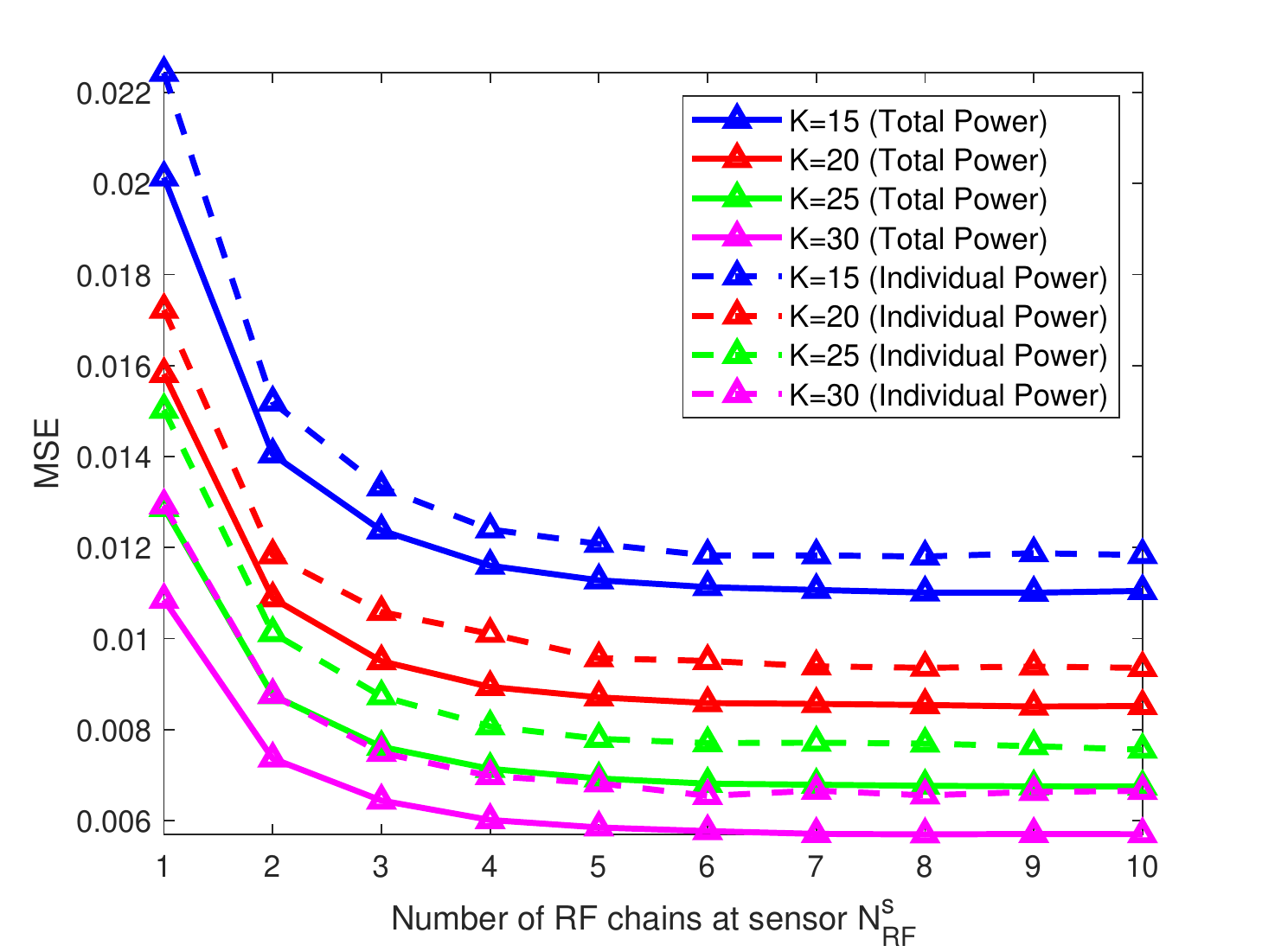}
\caption{ MSE versus RF chains $N_{\text{RF}}^s$ at each sensor for arbitrary-SNR sensor observations}
\label{noisy_mse_nrf}
\end{figure}

Fig. \ref{noisy_mse_nrf} portrays the MSE performance of the different designs derived in Section IV against the number of RF chains $N_{\text{RF}}^s$ at each sensor in the hybrid MIMO WSN for different numbers of sensors $K$. The MSE is seen to decrease upon increasing the number of RF chains $N_{\text{RF}}^s$ at each sensor. However, the improvement in MSE performance tends to become insignificant at high $N_{\text{RF}}^s$. This allows the sensors to have fewer RF chains, thereby reducing the power consumption and improving the battery life. This observation is applicable for both the total and individual sensor power constraints.
\section{Conclusions}
Hybrid transceiver designs have been developed for linear decentralized vector parameter estimation in mmWave MIMO WSNs for 5G-IoT applications. Optimal MSE transceivers were developed under both total as well as individual sensor power constraints. The centralized MMSE bound and BCRB were also determined for benchmarking the performance of the proposed hybrid transceiver designs. Simulation results demonstrated that the MSE performance of the proposed designs is close to that of the unconstrained fully-digital designs. Moreover, the proposed hybrid designs are also seen to perform better than the conventional beam-training schemes such as dominant directional precoding/combining. Future works can consider wireless powered sensor networks in IoT that can improve the battery lifetime of the sensor nodes.
\begin{appendices}
\section{Optimality of fully-digital precoder design} \label{maj}
A real-valued function $\textit{f}$ defined on a set $\mathcal{A}\subseteq{\mathbb{R}}^n$ is said to be Schur-concave on $\mathcal{A}$ if $\mathbf{x} \succ \mathbf{y} \Rightarrow \textit{f}\left(\mathbf{x}\right)\geq\textit{f}\left(\mathbf{y}\right)$. From majorization theory \cite{marshall1979inequalities}, a Schur-concave function $f$ is lower bounded as  $\textit{f}\left[\mathbf{\mathrm{diag}\left(\mathbf{E}\right)}\right]\geq \textit{f}\left[\mathbf{\boldsymbol{\lambda}\left(\mathbf{E}\right)}\right],$ with the lower bound achieved, when the matrix $\mathbf{E}$ is diagonal. Given that the trace of the error covariance matrix $\mathbf{E}$ is a Schur-concave function, the MSE reaches its lower bound, when the matrix $\mathbf{E}$ is diagonalized. Therefore, the fully-digital TPC designs in Section \ref{SectionIVA} and Section \ref{SectionIVB} are optimal since the resultant error covariance matrix which corresponds to the observation and FC noises, is diagonal.
\section{Proof of Theorem 1} \label{TPCproof}
The optimal fully-digital TPC designed in Section-III A under a total power budget was formulated in \eqref{E55} as
\begin{align*}
\mathbf{F}_k=\mathbf{V}_{g,k}^1\boldsymbol{\Sigma}\left(\mathbf{U}_{M,k}\right)^{\dagger}.
\end{align*}
From above, one can infer that the linear combination of the columns of $\mathbf{V}_{g,k}^1$ forms the columns of $\mathbf{F}_k$, which is expressed as
\begin{equation} 
\mathcal{C}(\mathbf{F}_k) \subseteq \mathcal{C}(\mathbf{V}_{g,k}^1). \label{eq:C1}
\end{equation}
Furthermore, since $\mathbf{V}_{g,k}^1$ is comprised of the $m$ left-most columns of $\mathbf{V}_{g,k}$, this implies
\begin{equation} \label{eq:C2}
\mathcal{C}(\mathbf{V}_{g,k}^1) \subseteq \mathcal{C}(\mathbf{V}_{g,k}).
\end{equation}  
The eigenvalue decomposition of $\mathbf{G}^H\mathbf{G}$ can be written as
\begin{equation} \label{eq:C3}
[\mathbf{G}_1,\mathbf{G}_2,\cdots,\mathbf{G}_K]^H \mathbf{G} = [\mathbf{V}_{g,1}^T,\mathbf{V}_{g,2}^T,\cdots,\mathbf{V}_{g,K}^T]^T\mathbf{\Lambda}_g \mathbf{V}_g^H.
\end{equation}
Multiplying both sides by the unitary matrix $ \mathbf{V}_g$ one obtains,
\begin{align}
& \mathbf{G}^H\mathbf{G}  \mathbf{V}_g=\mathbf{V}_{g}\mathbf{\Lambda}_g  \Rightarrow \mathbf{V}_{g}=\mathbf{G}^H\mathbf{G} \mathbf{V}_{g}\mathbf{\Lambda}_g^{\dagger}.
\end{align}
From the structure of \eqref{eq:C3}, one can infer that
\begin{align}
 \mathbf{V}_{g,k}=\mathbf{G}_k^H\mathbf{G} \mathbf{V}_{g}\mathbf{\Lambda}_g^{\dagger}.
\end{align}
From the previous argument, one can say that the column-space of $\mathbf{V}_{g,k}$ lies in the row-space of the mmWave MIMO channel matrix $\mathbf{G}_k$ for the $k$th sensor, i.e., we have
\begin{align}
\mathcal{C}(\mathbf{V}_{g,k}) & \subseteq \mathcal{C}(\mathbf{G}_k^H) = \mathcal{R}(\mathbf{G}_k). \label{eq:C4}
\end{align}
From \eqref{eq:C1}, \eqref{eq:C2}, \eqref{eq:C4} and \eqref{E36}, it follows that $\mathcal{C}(\mathbf{F}_k) \subseteq \mathcal{C}(\mathbf{V}_{g,k}^1) \subseteq \mathcal{C}(\mathbf{V}_{g,k})
\subseteq \mathcal{R}(\mathbf{G}_k) \subseteq \mathcal{C}(\mathbf{A}_{T,k})$, which proves the desired result.

\section{Hybrid TPC design for High-SNR/ Noiseless Sensor Observations} \label{noiseless}
In the high-SNR scenario \cite{6894624}, \cite{6905740}, where the observation
noise vector obeys $\mathbf{n} = 0$, the received signal of \eqref{eq6} is given by 
\begin{align}\label{eq 1}
\mathbf{y}={ \mathbf{G} }{\mathbf{F}}{ \mathbf{M} } \boldsymbol{\theta} + { \mathbf{v} }.
\end{align}
formulated as,
The MSE in the parameter estimation can be expressed as
\begin{align}  \label{MSE_noiseless}
\text{MSE}&=\text{Tr}\left(\left( \mathbf{I}_m + \frac{1}{\sigma_v^2}\mathbf{{M}}^H\mathbf{F}^H\mathbf{G}^H\mathbf{G}\mathbf{F}\mathbf{{M}}\right )^{-1}\right).
\end{align} 
\subsection{Hybrid TPC design under total power budget} The optimization
problem of the fully-digital TPC $\mathbf{F}$ is formulated as follows
\begin{equation} \label{eq:eq19}
\begin{aligned}
& \underset{\mathbf{{F}}}{\text{minimize}}
& &   \text{Tr}\left( \left(\mathbf{I}_m + \frac{1}{\sigma_v^2}\mathbf{M}^H\mathbf{{F}}^H\mathbf{G}^H\mathbf{G}\mathbf{{F}}\mathbf{M} \right)^{-1}\right)    \\
&\text{subject to}
& & \text{Tr}  \left( \mathbf{{F}}\mathbf{M}  \mathbf{{M}}^H\mathbf{F}^H \right) \le {P}_{T}.  \\
\end{aligned}
\end{equation}
Denoting $\mathbf{C} \triangleq \mathbf{{F}}\mathbf{M} \in \mathbb{C}^{KN_T \times m}$, the fully-digital TPC structure can be subsequently written as \cite{palomar2007mimo}
\begin{equation} \label{opt_maj}
\mathbf{{F}}\mathbf{M}=\mathbf{V}^1_g\mathbf{\Sigma}. 
\end{equation}
The MSE expression from \eqref{MSE_noiseless} can thus be rewritten as,
\begin{align} \label{scalar_obj_noiseless}
\text{MSE}&=\text{Tr}\left( \mathbf{I}_m + \frac{1}{\sigma_v^2}\mathbf{\Sigma}^H\mathbf{{\Lambda}}_g\mathbf{\Sigma}\right)^{-1} \nonumber \\
&=\sum\limits_{l=1}^m \frac{\sigma_v^2}{\sigma_v^2 + p_l\sigma^2_l({\mathbf{{G}}})}.
\end{align}
The total transmit power at high SNR is given by
  \begin{align} \label{scalar_cons_noiseless}
  \text{Tr}\left(\mathbf{{F}}\mathbf{M}\mathbf{{M}}^H \mathbf{F}^H\right)&=\text{Tr}\left(\mathbf{V}_g^1\mathbf{\Sigma}\mathbf{\Sigma}^H(\mathbf{V}^1_g)^H\right)=\sum\limits_{l=1}^m p_l.
  \end{align}
 Thus, the pertinent optimization problem can be reformulated using the objective in \eqref{scalar_obj_noiseless} and the constraint in \eqref{scalar_cons_noiseless} as
\begin{align}
 \underset{\mathbf{p}}{\text{minimize}}
\  &\sum\limits_{l=1}^m \frac{\sigma_v^2}{\sigma_v^2 + p_l\sigma^2_l(\mathbf{{G}})} \nonumber\\
\text{subject to}
\   &\sum\limits_{l=1}^m p_l \le P_T, \nonumber\\
& p_l \geq 0, 1 \leq l \leq m.
    \end{align}
Upon using the KKT conditions, the optimal value of $p_l$ is obtained as
 \begin{align}
 p_l= \left( \gamma\sqrt{\frac{\sigma_v^2}{\sigma_l^2(\mathbf{{G}})}} - \frac{\sigma_v^2}{\sigma_l^2(\mathbf{{G}})} \right) ^{+},
 \end{align}
where the Lagrangian multiplier $\gamma$ is given by:
 \begin{equation}
 \gamma= \frac{P_T + \sum\limits_{l=1}^m \frac{\sigma_v^2}{\sigma_l^2(\mathbf{{G}})}}{\sum\limits_{l=1}^m \sqrt{\frac{\sigma_v^2}{\sigma_l^2(\mathbf{{G}})}}}.
 \end{equation}
The optimal values $p_l, 1 \leq l \leq m$, upon substitution into \eqref{opt_maj} yield the matrix 
 $ \mathbf{{F}}\mathbf{M}$. The individual fully-digital TPCs $\mathbf{F}_{k}$, $1 \leq k \leq K$, can be expressed as
\begin{equation} \label{E33}
\mathbf{F}_k= \mathbf{V}^1_{g,k}\mathbf{\Sigma} \left(\mathbf{{M}}_k\right)^{\dagger},
\end{equation}
where $\mathbf{V}^1_{g,k}$ denotes the sub-matrix of $\mathbf{V}^1_{g}$ comprised of the rows $(k-1)N_T + 1$ to $kN_T$ and all the columns. Now, given the optimal fully-digital TPC $\mathbf{F}_{k}$ of the $k$th sensor, its components $\mathbf{F}_{\text{RF},k}$ and $\mathbf{F}_{\text{BB},k}$ can be designed by exploiting the SOMP algorithm.\par
\subsection{Hybrid TPC design under individual sensor power constraints}Using majorization theory, the hybrid TPC structure in \eqref{opt_maj} is valid
also for this system, which can be subsequently written as
\begin{align} 
&\mathbf{{F}}\mathbf{M}=\mathbf{V}^1_g\mathbf{\Sigma} \nonumber \\ &\Rightarrow \begin{bmatrix}
\mathbf{F}_1 & \mathbf{0}& \cdots &\mathbf{0}\\
\mathbf{0}& \mathbf{F}_2 & \cdots & \mathbf{0}\\
\vdots& & \ddots & \vdots\\
\mathbf{0}&\mathbf{0} &\cdots & \mathbf{F}_K
\end{bmatrix}\begin{bmatrix}
\mathbf{M}_{1} \\
\mathbf{M}_{2}\\
\vdots\\
\mathbf{M}_{K}
\end{bmatrix} =\begin{bmatrix}
\mathbf{V}_{g,1}^1 \\
\mathbf{V}_{g,2}^1\\
\vdots\\
\mathbf{V}_{g,K}^1\end{bmatrix}\mathbf{\Sigma} \nonumber \\
&\Rightarrow \mathbf{F}_k\mathbf{M}_{k}=\mathbf{V}_{g,k}^1\mathbf{\Sigma}, \ \forall k. 
\end{align}
Substituting the above equation into the individual sensor power constraint, one obtains
\begin{align}
\text{Tr}&\left( \mathbf{F}_k  \mathbf{M}_k  \mathbf{M}_k^H  { \mathbf{F} }^{ H }_k\right)=\text{Tr}\left( \mathbf{V}_{g,k}^1\mathbf{\Sigma}  \mathbf{\Sigma}^H  \left(\mathbf{V}_{g,k}^1\right)^{ H }\right) \nonumber \\
& =\text{Tr}\left(\underbrace{(\mathbf{V}_{g,k}^1)^{ H } \mathbf{V}_
{g,k}^1}_{\boldsymbol{\Phi}_k}  \boldsymbol{\Sigma}\boldsymbol{\Sigma}^H\right)  = \sum_{l=1}^m p_l\left[\boldsymbol{\Phi}_k\right]_{ll}.
\end{align}
 Hence, the scalar-valued optimization problem of minimizing the MSE of the estimate of the parameter vector $\boldsymbol{\theta}$ can be equivalently formulated as
\begin{align} \label{E67}
 \underset{\mathbf{p}}{\text{minimize}}
\  &\sum\limits_{l=1}^m \frac{\sigma_v^2}{\sigma_v^2 + p_l\sigma^2_l(\mathbf{{G}})} \nonumber\\
\text{subject to}
\   &\sum_{l=1}^m p_l
\left[\boldsymbol{\Phi}_k\right]_{ll} \le P_k\ \forall k, \nonumber\\
& p_l \geq 0, 1 \leq l \leq m. 
    \end{align}
Subsequently, the optimal $\mathbf{p}$ can be found using suitable convex solvers. The fully-digital optimal TPC $\mathbf{F}_k$ at $k$th sensor is obtained from \eqref{E33} as $\mathbf{F}_k= \mathbf{V}^1_{g,k}\mathbf{\Sigma} \left(\mathbf{{M}}_k\right)^{\dagger}$. The corresponding baseband and RF components of the
fully-digital hybrid TPC are then designed using Algorithm
1.

\section{Equivalence of \eqref{eq182} and \eqref{eq40}} \label{RCproof}
The cost function in \eqref{eq182} may be rewritten as
\begin{align} \label{eq99}
 &\mathbb{E}\left \{ \left\Vert \boldsymbol{\theta}-\mathbf{W}_{\text{BB}}^{H}\mathbf{W}_{\text{RF}}^{H}\mathbf{y} \right\Vert_2^2\right \}   \nonumber \\
 &=\mathbb{E}\left\{\text{Tr}\left( ({\mathbf{ \boldsymbol{\theta}}}-{\mathbf{W}}_{\text{BB}}^{H}{\mathbf{W}}_{\text{RF}}^{H}{\mathbf{y}}) ({\mathbf{ \boldsymbol{\theta}}}-{\mathbf{W}}_{\text{BB}}^{H}{\mathbf{W}}_{\text{RF}}^{H}{\mathbf{y}})^{H}\right)\right\} \nonumber \\
& = \text{Tr}\left(\mathbb{E}\left\{ \boldsymbol{\theta}\mathbf{ \boldsymbol{\theta}}^{H} \right\}\right) -2\Re\left\lbrace \text{Tr}\left( \mathbb{E}\left\{\mathbf{ \boldsymbol{\theta}}\mathbf{y}^{H}\right\}\mathbf{W}_{\text{RF}}\mathbf{W}_{\text{BB}} \right) \right\rbrace  \nonumber \\
&+ {\Tr}\left(\mathbf{W}_{\text{BB}}^{H}\mathbf{W}_{\text{RF}}^{H} \mathbb{E}\left\{ \mathbf{y}\mathbf{y}^{H}\right\} \mathbf{W}_{\text{RF}}\mathbf{W}_{\text{BB}}  \right),
\end{align}
where $\Re(.)$ denotes the real part of the complex value. Note that the optimization variables are $\mathbf{W}_{\text{RF}}$ and $\mathbf{W}_{\text{BB}}$. Therefore, the constant terms ${\Tr}\left(\mathbf{W}^{H}\mathbb{E}\left\{ \mathbf{y}\mathbf{y}^{H}\right\}\mathbf{W}\right) - \text{Tr}\left(\mathbb{E}\left\{\boldsymbol{\theta}\boldsymbol{\theta}^{H}\right\}\right)$ are added to the cost function in \eqref{eq99}, where $\mathbf{W}$ is the digital LMMSE combiner. The subsequent steps are as follows
\begin{align} \label{eq15}
&\mathbb{E}\left \{ \left\Vert \boldsymbol{\theta}-\mathbf{W}_{\text{BB}}^{H}\mathbf{W}_{\text{RF}}^{H}\mathbf{y} \right\Vert_2^2\right \} \nonumber \\
& ={\Tr}\left(\mathbf{W}^{H}\mathbb{E}\left\{\mathbf{y}\mathbf{y}^{H}\right\}\mathbf{W}\right) -2\Re\left\lbrace \text{Tr}\left( \mathbb{E}\left\{\mathbf{\boldsymbol{\theta}}\mathbf{y}^{H}\right\}\mathbf{W}_{\text{RF}}\mathbf{W}_{\text{BB}} \right) \right\rbrace\nonumber \\& \ \
 + \text{Tr}\left(\mathbf{W}_{\text{BB}}^{H}\mathbf{W}_{\text{RF}}^{H} \mathbb{E}\left\{ \mathbf{y}\mathbf{y}^{H}\right\}\mathbf{W}_{\text{RF}}\mathbf{W}_{\text{BB}}  \right) \nonumber \\
 & \stackrel{a}{=} \text{Tr}\left(\mathbf{W}^{H}{\mathbf{R}_{yy}}\mathbf{W}\right) -2\Re\left\lbrace \text{Tr}\left( \mathbf{W}^{H}\mathbf{R}_{yy}\mathbf{W}_{\text{RF}}\mathbf{W}_{\text{BB}} \right) \right\rbrace\nonumber \\
& + \text{Tr}\left(\mathbf{W}_{\text{BB}}^{H}\mathbf{W}_{\text{RF}}^{H} \mathbf{R}_{yy} \mathbf{W}_{\text{RF}}\mathbf{W}_{\text{BB}}  \right) \nonumber \\
&= \text{Tr}\left(\left( \mathbf{W}^{H}-\mathbf{W}_{\text{BB}}^{H}\mathbf{W}_{\text{RF}}^{H}\right) \mathbf{R}_{yy}\left( \mathbf{W}^{H}-\mathbf{W}_{\text{BB}}^{H}\mathbf{W}_{\text{RF}}^{H}\right)^{H}\right) \nonumber \\
& =  \left\Vert {\mathbf{R}_{yy}^{\frac{1}{2}}\left(\mathbf{W}-\mathbf{W}_{\text{BB}}\mathbf{W}_{\text{RF}}\right)}\right\Vert_{F}^{2},
\end{align}
where (a) follows by modifying the second term as 
\begin{align*}
\text{Tr}\left( \mathbb{E}\left\{\mathbf{\boldsymbol{\theta}}\mathbf{y}^{H}\right\}\mathbf{W}_{\text{RF}}\mathbf{W}_{\text{BB}} \right) &=\text{Tr}\left( \mathbb{E}\left\{\mathbf{\boldsymbol{\theta}}\mathbf{y}^{H}\right\}\mathbf{R}_{yy}^{-1}\mathbf{R}_{yy}\mathbf{W}_{\text{RF}}\mathbf{W}_{\text{BB}} \right) \nonumber \\
&=\text{Tr}\left(\mathbf{W}^{H}\mathbf{R}_{yy}\mathbf{W}_{\text{RF}}\mathbf{W}_{\text{BB}} \right).
\end{align*}
The last step in the above expression is obtained from \eqref{eq44} by substituting $\mathbb{E}\left\{\mathbf{\boldsymbol{\theta}}\mathbf{y}^{H}\right\}\mathbf{R}_{yy}^{-1}$ with $\mathbf{W}^H$.
\end{appendices}
\vspace{-10pt}
\bibliographystyle{IEEEtran}
\bibliography{mmWave_WSN_manuscript}
\begin{IEEEbiography}
[{\includegraphics[width=1in,height=1.25in,clip,keepaspectratio]{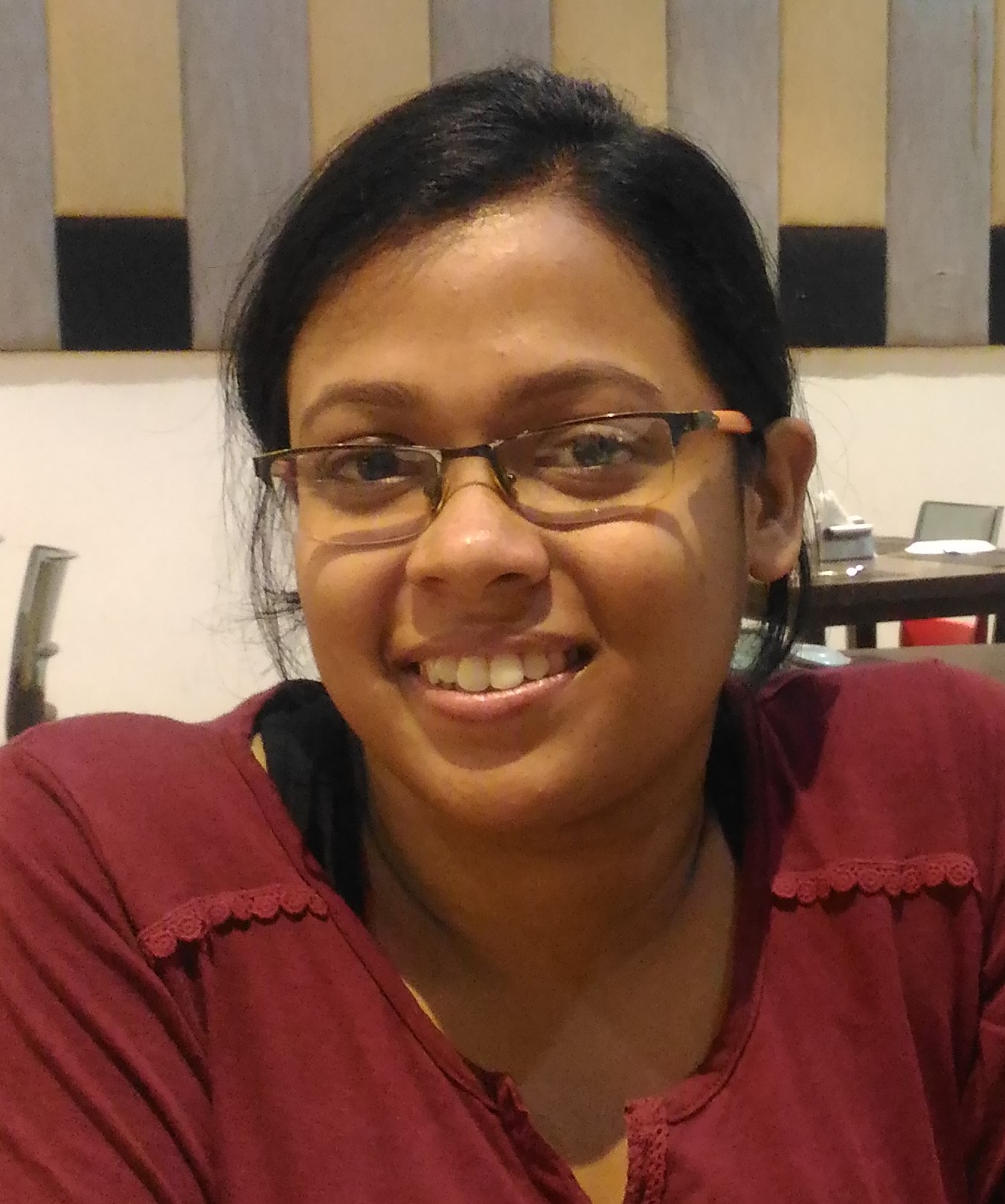}}]{Priyanka Maity} received the B.Tech. degree
in electronics and communication engineering
from the Kalyani Government Engineering
College, West Bengal University of Technology,
India, in 2016 and M.Tech. degree in communication
and networks from NIT Rourkela in 2018.
She is currently working towards the Ph.D. degree
with the Department of Electrical Engineering, IIT
Kanpur, Kanpur, India. Her research interests include
wireless communications, signal processing
techniques for wireless communication, MIMO, millimeter wave communication,
Wireless Sensor Networks for IoT. She was a finalist for the Qualcomm Innovation Fellowship in 2022.
\end{IEEEbiography}

\begin{IEEEbiography}
[{\includegraphics[width=1in,height=1.25in,clip,keepaspectratio]{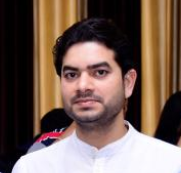}}]{Suraj Srivastava} received the M.Tech. degree in Electronics and Communication Engineering from Indian Institute of Technology Roorkee, India, in 2012, and Ph.D. degree in Electrical Engineering from Indian Institute of Technology Kanpur, Kanpur, India, in 2022. From July 2012 to November 2013, he was employed as a Staff-I systems design engineer with Broadcom Research India Pvt. Ltd., Bangalore, and from November 2013 to December 2015, he was employed as a lead engineer with Samsung Research India, Bangalore where he worked on developing layer-2 of the 3G UMTS/WCDMA/HSDPA modem. His research interests include applications of Sparse Signal Processing in 5G Wireless Systems, mmWave and Tera-Hertz Communication, Orthogonal Time-Frequency Space (OTFS), Joint Radar and Communication (RadCom), Optimization and Machine Learning. He was awarded Qualcomm Innovation Fellowship (QIF) in year 2018 and 2022 from Qualcomm. He was also awarded Outstanding Ph.D. Thesis and Outstanding Teaching Assistant awards from IIT Kanpur.
\end{IEEEbiography}

\begin{IEEEbiography}
[{\includegraphics[width=1in,height=1.25in,clip,keepaspectratio]{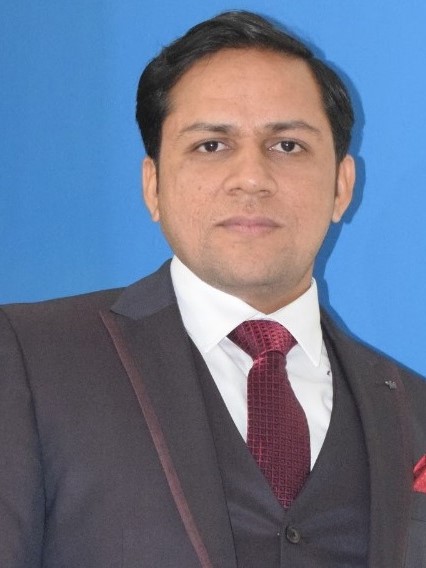}}]{Kunwar Pritiraj Rajput}  received the B.E. degree in electronics and communication engineering from Jabalpur Engineering
College, Jabalpur, India, in 2010; the M.Tech. degree in digital communication from the Atal Bihari Vajpayee Indian Institute of Information Technology and Management, Gwalior, India, in 2013; and the Ph.D. degree in 2022 from IIT Kanpur, Kanpur, India. He is currently working as a research associate in the SPARC group at SnT, University of Luxembourg. His research interests include decentralized and distributed parameter estimation in MIMO, massive MIMO, mmWave MIMO wireless sensor networks, sparse signal processing and ISAC. He worked as Vice Chair of the IEEE Signal Processing Society, Student Chapter Branch, IIT Kanpur. He was given the best teaching assistant (TA) award for the 5G wireless communication course at IIT Kanpur. He was also a finalist for the Qualcomm Innovation Fellowship in 2022.
\end{IEEEbiography}

\begin{IEEEbiography}
[{\includegraphics[width=1in,height=1.25in,clip,keepaspectratio]{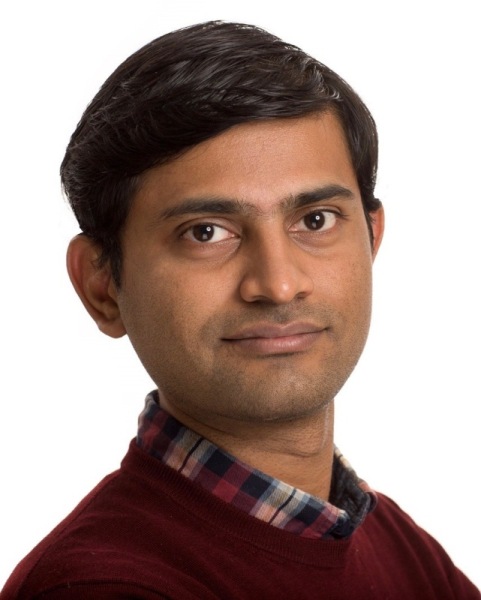}}]{Naveen K. D. Venkategowda} received the B.E. degree in electronics and communication engineering from Bangalore University, Bengaluru, India, in 2008, and the Ph.D.  degree in electrical engineering from Indian Institute of Technology, Kanpur, India, in 2016. He is currently an Universitetslektor at the Department of Science and
Technology, Linköping University, Sweden.  From Oct. 2017 to Feb. 2021, he was postdoctoral researcher at the Department of Electronic Systems, Norwegian University of Science and Technology, Trondheim, Norway. He was a Research Professor at the School of Electrical Engineering, Korea University, South Korea from Aug. 2016 to Sep. 2017. He was a recipient of the TCS Research Fellowship (2011-15) from TCS for graduate studies in
computing sciences, the ERCIM Alain Bensoussan Fellowship in 2017, and Runner up - Best Paper Award in Fusion 2020.
\end{IEEEbiography}

\begin{IEEEbiography}
[{\includegraphics[width=1.30in,height=1.25in,clip,keepaspectratio, angle=90]{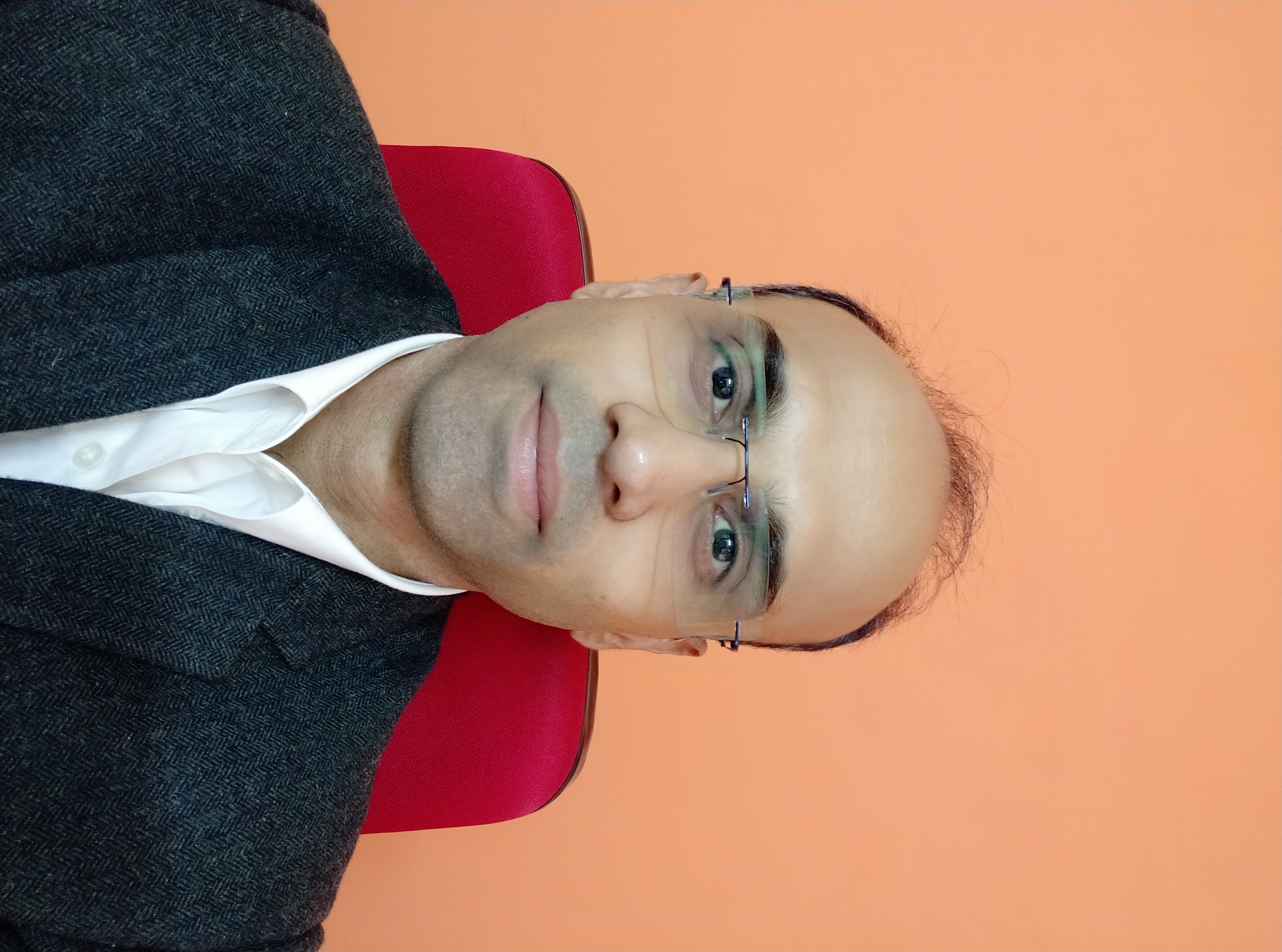}}]{Aditya K. Jagannatham (Senior Member, IEEE)} received the bachelor’s degree from the Indian Institute of Technology, Bombay, and the M.S. and Ph.D. degrees from the University of California, San Diego, CA, USA. From April 2007 to May 2009, he was employed as a Senior Wireless Systems Engineer with Qualcomm Inc., San Diego, CA where he was a part of the Qualcomm CDMA Technologies (QCT) Division. His research interests include next-generation wireless cellular and WiFi networks, with a special emphasis on various 5G technologies, such as massive MIMO, mmWave MIMO, FBMC, NOMA, full duplex, and others. He is currently a Professor with the Electrical Engineering Department, IIT Kanpur, where he also holds the Arun Kumar Chair Professorship. He has been twice awarded the P. K. Kelkar Young Faculty Research Fellowship for excellence in research, the Qualcomm Innovation Fellowship (QInF), the IIT Kanpur Excellence in Teaching Award, the CAL(IT)2 fellowship at the University of California San Diego and the Upendra Patel Achievement Award at Qualcomm.
\end{IEEEbiography}

\begin{IEEEbiography}
[{\includegraphics[width=1.in,height=1.25in,clip,keepaspectratio]{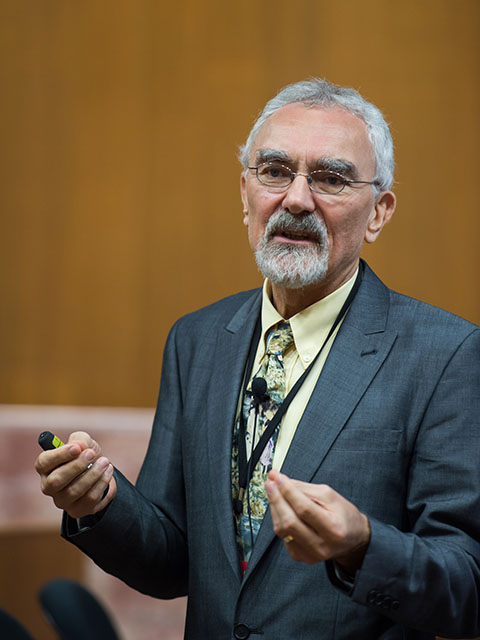}}]{Lajos Hanzo (Life Fellow, IEEE)} received his 5-year Master degree in electronics in 1976 and his doctorate in 1983.  He holds an honorary doctorate by the Technical University of Budapest (2009) and by the University of Edinburgh (2015).  He is a member of the Hungarian Academy of Sciences and a former Editor-in-Chief of the IEEE Press. He is a Governor of both IEEE ComSoc and of VTS. He has published 1900+ contributions at IEEE Xplore and supervised 119 PhD students.
\end{IEEEbiography}
\end{document}